\documentclass[12pt]{article}
\usepackage{amsfonts}
\usepackage{amsmath}
\usepackage{cite}
\usepackage[decmulti]{inputenc}

\input{tcilatex}

\begin{document}

\title{Classes of admissible exchange-correlation density functionals for
pure spin and angular momentum states}

\author{A.L. Tchougr\'{e}eff \\
JARA, Institute of Inorganic Chemistry, \\
RWTH Aachen, Germany; \\
Poncelet Laboratory, \\
Independent University of Moscow,\\
Moscow Center for Contunous \\
Mathematical Education; \\
Division of Electrochemistry, Department of Chemistry, \\
Moscow State University, 119991, Moscow, Russia \\
 and \\
J.G. \'{A}ngy\'{a}n \\
CRM2,  Nancy-University, CNRS, \\ B.P.~239, Vandoeuvre-l\`{e}s-Nancy, France}
\maketitle

\newpage
\begin{abstract}
We analyze the various approaches to construct exchange-correla\-tion
functionals which are able to describe states of definite spin multipli\-city
in the DFT realm and outline the characteristics of possible functionals
consistent with the Kohn-Sham theory. To achieve this goal the unitary group
technique is applied to label many-electron states of definite total spin
and to calculate the corresponding analogs of the Roothaan coupling
coefficients. The possibility of using range separated  Coulomb potential of electron-electron interaction for constructing functionals discriminating multiplet states in the $d$-shells is explored and a tentative system of
state-specific functionals, covering nontrivial correlations in  $d$-shells of transition metal ions, is proposed for the Fe$^{2+}$ ions.
\end{abstract}

\section{Introduction}

Although the Density Functional Theory (DFT) based methods of modeling
electronic structure of molecules and solids widely proliferate during last
decades \cite{Parr:94:book,March:book,Koch:Holthausen}, the problem of consistent
description of transition metal and rare earth compounds with open $d$- and $%
f$-shells, respectively, remains a still unresolved, challenging problem in
this framework \cite{Darkhovskii:06}. One of the main reasons for this failure of
the DFT is that the multiplet spin/orbital momentum states are generally not
easily described within the DFT paradigm. The source of that intimate
"unfriendlieness" of the DFT to the multiplet states lays in the
"oversymmetry" of the fundamental quantity pertaining in the realm of DFT: 
the one-electron density. As it has been demonstrated many times, states of
different total spin and/or spatial symmetry may produce equal one-electron
densities. The complication arising from this is the impossibility to distinguish  the nature of the ground state on the basis of the total density only: although only one of say two functions represents the ground state, i.e.\
the \emph{exact} energies of the involved states may be different, they 
turn out to be degenerate in the DFT context. In other terms, if the same densities are fed to the "universal" density functional implied by the 
DFT, it is going to produce the same value of the electronic energy for 
states whose \emph{exact} energies are \emph{different}. 
Of course the latter remark may be opposed by noting that the "universal" functional is going to output the ground state energy only, but in this case it is not clear how other important pieces of information concerning the nature of this ground state (e.g.\ its spin multiplicity) can be extracted from such an answer. 
 
This situation certainly requires some clarification which is addressed in the present paper. In order to do so we give below a brief description of relevant
elements of the electronic structure theory (Section~\ref{Theory}). Then we
consider an archetypical example of the problems encountered by the DFT while
trying to reproduce correct spin properties of many-electron systems
(Section~\ref{Archetyp}). Then we propose a general scheme allowing to
include states of definite spin in the DFT theory (Section~\ref{Unitary}).
This, however, does not solve the problem of the multiple states in the open 
$d$- and $f$-shells of the transition and rare earth ions. For this end we
explore in Section \ref{Short-Long} the possibility to circumvent these
problems with use of the short/long range separation of Coulomb interaction
between electrons and propose in Section~\ref{d-shell} some conceivable
state-dependent definition of exchange-correlation functionals capable to
reproduce the energies of nontrivially correlated many-electronic states in
the $d$-shell of the Fe$^{2+}$ ions.

\section{Theoretical background\label{Theory}}

\subsection{Electronic distribution.}

The main idea of the DFT is to reduce the description of entire electronic
structure to a single quantity: the one-electron density -- the diagonal
part of the one-electron density matrix.  The possibility of such a reduction is
proven by the Hohenberg-Kohn theorems \cite{HK:64} which state
an existence of a universal one-to-one correspondence between one-electron
external potential and the one-electron density in that sense that not only
the one-electron potential acting upon a given number of electrons uniquely
defines the ground state of such a system \textit{i.e.} its wave function
and thus the one-electron density, but also that for each given density
integrating to a given number of electrons $N$ a one-electron potential
yielding that given density can be uniquely defined from the density. 
The  "density only" formulation of the electronic
structure problem, even if it is practically achieved, leaves unanswered an
important question of the nature of the ground state thus obtained \textit{%
e.g.} about its total spin (or other symmetry features).

Incidentally, the symmetry properties of quantum states, like total spin,
are easier expressed in terms of wave functions (see Ref. \cite{Blokhintsev:book}) 
so it would be practical to consider tentative relation between the wave
function and density only pictures of the electronic structure. The required
relation can be established with use of the reduced one- and two-electron
density \emph{matrices} as much simpler objects than the wave functions,
providing equivalent description of electronic structure. The reduced
density matrices respectively depend on two ($x,x^{\prime }$) and four ($%
x_{1}x_{2},x_{1}^{\prime }x_{2}^{\prime }$) coordinates: 
\begin{equation}
\begin{array}{lll}
\displaystyle\rho ^{(1)}(x;x^{\prime }) & = & N\int \Psi ^{\ast
}(x,x_{2},\ldots x_{N})\times \\ 
\displaystyle & \times & \Psi (x^{\prime },x_{2},\ldots ,x_{N})dx_{2}\ldots
dx_{N}, \\ 
\displaystyle\rho ^{(2)}(x_{1}x_{2};x_{1}^{\prime }x_{2}^{\prime }) & = & 
\frac{N(N-1)}{2}\int \Psi ^{\ast }(x_{1},x_{2},x_{3},\ldots x_{N})\times \\ 
\displaystyle & \times & \Psi (x_{1}^{\prime },x_{2}^{\prime },x_{3},\ldots
,x_{N})dx_{3}\ldots dx_{N},%
\end{array}
\label{DensityMatrices}
\end{equation}%
where the composite electronic coordinate $x$ represents a pair $\left( 
\mathbf{r},s\right) $ of three dimensional radius vector $\mathbf{r}$ of an
electron and of the discrete variable $s$ taking either of the two allowed
values $\pm \hbar /2$ of the projection of electronic spin. The transition
to the description in terms of reduced density matrices is itself a
significant simplification (although being absolutely exact). The
one-electron density implied by the DFT theory appears then as a result of
further reduction of eq. (\ref{DensityMatrices}):%
\begin{equation}
\rho (\mathbf{r})=\sum\limits_{s}\rho ^{(1)}\mathbf{({r}}s\mathbf{;{r}}s%
\mathbf{)}  \label{Density}
\end{equation}%
Thus according to the DFT paradigm the one-electron density which depends on
one spatial radius-vector must be able to serve as an equivalent substitute
to the $N$-electronic wave function dependent on $N$\ radius vectors and $N$
more spin projections of all electrons involved. The obvious loss of
information which takes place by going from eq. (\ref{DensityMatrices}) to
eq. (\ref{Density}) -- we remind that going from the wave function $\Psi
(x_{1},x_{2},x_{3},\ldots x_{N})$ to the reduced density matrices by eq. (%
\ref{DensityMatrices}) does not produce any loss of at least important
information -- must be compensated by the \ "universal" and \ "exact"
density functional which is in general unknown.

\subsection{Electronic energy.}

Leaving aside the  "ideal" DFT using the unknown "universal" and
"exact" functional of the density eq. (\ref{Density}) and turning to
pragmatic methods pertaining to the DFT realm needs some approximate
expressions for the energy presented as a functional of the density eq. (\ref%
{Density}) only. In the wave function and in the equivalent reduced density
matrix formulations the energy has the form: 
\begin{equation}
\begin{array}{lll}
E & = & \langle \hat{T}_{e}\rangle +\langle \hat{V}_{ne}(\left\{ \mathbf{R}%
\right\} )\rangle +\langle \hat{V}_{ee}\rangle +V_{nn}(\left\{ \mathbf{R}%
\right\} ).%
\end{array}
\label{Energy}
\end{equation}%
where $\left\{ \mathbf{R}\right\} $\ stands for the set of radius-vectors of
all nuclei inducing the electrostatic potential external to the electrons of
the system. In the coordinate representation the above averages acquire
familiar appearance: 
\begin{equation}
\begin{array}{l}
\left\langle \hat{T}_{e}\right\rangle =-\dfrac{1}{2}\sum\limits_{s}\dint%
\limits_{\mathbf{r}=\mathbf{r}^{\prime }}\Delta ^{\prime }\rho ^{(1)}(%
\mathbf{r}s;\mathbf{r}^{\prime }s)d\mathbf{r} \\ 
\left\langle \hat{V}_{ne}(\left\{ \mathbf{R}\right\} )\right\rangle
=\dsum\limits_{i}Z_{i}\dint \dfrac{\rho (\mathbf{r})d\mathbf{r}}{\left\vert 
\mathbf{R}_{i}-\mathbf{r}\right\vert } \\ 
\left\langle \hat{V}_{ee}\right\rangle =\dfrac{1}{2}\sum\limits_{ss^{\prime
}}\dint \dint \dfrac{\rho ^{(2)}(\mathbf{r}s,\mathbf{r}^{\prime }s^{\prime };%
\mathbf{r}s,\mathbf{r}^{\prime }s^{\prime })}{\left\vert \mathbf{r}-\mathbf{r%
}^{\prime }\right\vert }d\mathbf{r}d\mathbf{r}^{\prime } \\ 
V_{nn}(\left\{ \mathbf{R}\right\} )=\dfrac{1}{2}\dsum\limits_{i\neq j}\dfrac{%
Z_{i}Z_{j}}{|\mathbf{R}_{i}-\mathbf{R}_{j}|};\mathrm{where} \\ 
\Delta ^{\prime }=\frac{\partial ^{2}}{\partial x^{\prime 2}}+\frac{\partial
^{2}}{\partial y^{\prime 2}}+\frac{\partial ^{2}}{\partial z^{\prime 2}}%
\end{array}
\label{EnergyComponents}
\end{equation}%
where the expressions eq. (\ref{EnergyComponents}) are assumed to be
specific for a given geometry $\left\{ \mathbf{R}\right\} $ and for an
electronic state described by the $N$-electronic wave function $\Psi =\Psi
\left( x_{1},...,x_{N}\right) $ employed to define the density matrices eq. (%
\ref{DensityMatrices}). The first row in eq. (\ref{EnergyComponents}) is the
kinetic energy of electrons, the second row is the energy of Coulomb
attraction of electrons to nuclei, the third row is the energy of
interelectronic repulsion; the fourth one is the energy of Coulomb repulsion
of the nuclei, which does not depend on the electronic density/wavefunction.

In the above expressions eqs~(\ref{Energy}), (\ref{EnergyComponents}) only
the average of the nuclear potential $\hat{V}_{ne}$ is \emph{exactly} a
functional of the required form: that of the one-electron density eq.~(\ref%
{Density}). All other terms in eqs~(\ref{Energy}), (\ref{EnergyComponents})
require further consideration. It applies similarly to the remaining\ one-
and two-electron contributions to the energy. As for the one-electron term,
the kinetic energy requires knowledge of the one-electron density \emph{%
matrix} eq.~(\ref{DensityMatrices}) rather than its diagonal part eq.~(\ref%
{Density}) although effectively in a narrow range of spatial separations $%
\mathbf{r}-\mathbf{r}^{\prime }$ which must be sufficient to determine the
second derivative. The attempts to avoid this bottleneck and to obtain
pragmatic DFT methods brought Kohn and Sham \cite{Kohn:65} to their
famous orbital construct which allowed them to express the kinetic energy in
terms of some single-determinant wave function yielding by definition the
required (exact) one-electron density. Then the kinetic energy is calculated
as one of the system of non-interacting electrons described by a single determinant  built of KS orbitals.

\subsection{Electronic density and electronic energy decompositions.}

While the Kohn-Sham construct offers an efficient technique to handle the
difficult kinetic energy problem and provide a very good first approximation
to it, the representation of the electron-electron interaction energy in
terms of the one-electron density (and possibly further parameters derived
from the KS determinant) remains the central problem on modern density
functional theory. 
Generally, calculating the Coulomb electron-electron energy (3-rd row of eq.
(\ref{EnergyComponents})) requires knowledge of the two-electron density
matrix. According to \cite{Loewdin:59} it decomposes: 
\begin{equation}
\rho ^{(2)}(x_{1},x_{2};x_{1}^{\prime },x_{2}^{\prime })=\frac{1}{2}%
\left\vert 
\begin{array}{cc}
\rho ^{(1)}(x_{1};x_{1}^{\prime }) & \rho ^{(1)}(x_{2};x_{1}^{\prime }) \\ 
\rho ^{(1)}(x_{1};x_{2}^{\prime }) & \rho ^{(1)}(x_{2};x_{2}^{\prime })%
\end{array}%
\right\vert -\chi (x_{1},x_{2};x_{1}^{\prime },x_{2}^{\prime }),
\label{2Density}
\end{equation}%
where the first (determinant) term corresponds to the part of the two-particle density matrix which can be accounted for even in the independent electrons
approximation. The second term in eq.~(\ref{2Density}) -- the cumulant of
the two-particle density matrix -- is responsible for deviation of electrons'
behavior from the independent electron model, \textit{i.e.} for their
Coulomb correlations. The Coulomb interaction of electrons eq.~(\ref%
{EnergyComponents}) can be decomposed to contributions associated to the
terms of the above two-particle density matrix decomposition eq. (\ref%
{2Density}): 
\begin{eqnarray}
\left\langle V_{ee}\right\rangle &=&E_{H}+\overline{E}_{xc};
\label{InteractionDecomposition} \\
\overline{E}_{xc} &=&\overline{E}_{x}+\overline{E}_{c}.  \notag
\end{eqnarray}%
by singling out first the \textquotedblleft classical\textquotedblright\
part of the Coulomb interaction energy -- the Hartree energy: 
\begin{equation}
E_{H}=\frac{1}{2}\sum\limits_{ss^{\prime }}\int \frac{\rho ^{(1)}(\mathbf{r}s%
\mathbf{,{r}}s\mathbf{)}\rho ^{(1)}\mathbf{({r}^{\prime }}s^{\prime }\mathbf{%
,{r}^{\prime }}s^{\prime }\mathbf{)}}{|\mathbf{{r}-{r}^{\prime }|}}d\mathbf{r%
}d\mathbf{{r}^{\prime }}=\frac{1}{2}\int \frac{\rho (\mathbf{r)}\rho \mathbf{%
({r}^{\prime })}}{|\mathbf{{r}-{r}^{\prime }|}}d\mathbf{r}d\mathbf{{r}%
^{\prime }}  \label{HartreeEnergy}
\end{equation}%
and then the exchange and correlation energies: 
\begin{eqnarray}
\overline{E}_{x} &=&-\frac{1}{2}\sum\limits_{s}\int \frac{\rho ^{(1)}(%
\mathbf{r}s\mathbf{,{r}^{\prime }}s\mathbf{)}\rho ^{(1)}\mathbf{({r}^{\prime
}}s\mathbf{,{r}}s\mathbf{)}}{|\mathbf{r}-\mathbf{r}^{\prime }|}d\mathbf{r}d%
\mathbf{{r}^{\prime }}  \label{ExchangePart} \\
\overline{E}_{c} &=&-\frac{1}{2}\sum\limits_{ss^{\prime }}\int \frac{\chi (%
\mathbf{r}s\mathbf{,\mathbf{{r}^{\prime }}}s\mathbf{^{\prime };\mathbf{r}}s%
\mathbf{,{r}^{\prime }}s^{\prime }\mathbf{)}}{|\mathbf{r}-\mathbf{r}^{\prime
}|}d\mathbf{r}d\mathbf{{r}^{\prime }}  \label{CorrelationPart}
\end{eqnarray}%
whose definitions eqs~(\ref{ExchangePart}), (\ref{CorrelationPart}) are
given respectively in terms of the off-diagonal part of one-electron density
matrix $\rho ^{(1)}\mathbf{({r}^{\prime }}s\mathbf{,{r}}s\mathbf{)}$ and of
the two-electron density matrix cumulant $\chi (\mathbf{r}s\mathbf{,\mathbf{{%
r}^{\prime }}}s^{\prime }\mathbf{\mathbf{;\mathbf{r}}}s\mathbf{,{r}^{\prime }%
}s^{\prime }\mathbf{)}$ -- the difference between the exact two-electron
density matrix and its Hartree-Fock (self-consistent field) estimate.

While the definition of the Hartree-energy is unique, and constitutes
together with the nuclear-electron repulsion energy the part of the total
energy that can be written straightforwardly as a simple analytic functional
of the one-particle density, the exchange and correlation energies are
defined in quantum chemistry and in DFT in different ways. As far as the
exchange energy is concerned, one should remark that the one-particle
density matrix, appearing in eq.~(\ref{ExchangePart}) is supposed to be
exact. This quantity is not available even in exact KS theory, where 
we have at best the one-particle density matrix associated to the single determinant constructed from the exact KS orbitals. By consequence, the exact exchange
energy in DFT is in general not equal to $\overline{E}_{x}$.

It must be observed that the usual definition of the correlation energy in
quantum chemistry, proposed by L\"{o}wdin in Ref. \cite{Loewdin:59} differs
from that given in Eq.~(\ref{CorrelationPart}), which follows rather the
suggestion due to Kutzelnigg and Mukherjee~\cite{Kutzelnigg:93,Kutzelnigg:03}. This latter definition has the conceptual advantage that it uses the
quantities entering eqs. (\ref{HartreeEnergy}) - (\ref{CorrelationPart})
irrespective to any approximate method of calculation of the electronic
energy. Some authors, (cf. \textit{e.g.} Refs. \cite{Fritsche:03,Ziesche:02})
argue that pragmatic DFT methods can be considered as 
approximations to the two-electron density matrix cumulant.

This situation is as well slightly more complicated in conventional
Kohn-Sham theory, where the correlation energy involves also the difference
of the exact and KS kinetic energies. However, this kinetic energy
contribution can be assimilated to a potential energy term the virtue of the
adiabatic connection procedure, which allows one to write the total correlation energy as an average electron-electron interaction over the adiabatic 
connection path.

\section{Symmetry non-sensitivity of the density-only methods \label{Archetyp}%
}

\subsection{Archetypical example of existing problems}

In order to better understand the problems which appear in the DFT realm
when trying to describe the correct total spin of a many electronic state we
consider the simplest system of \ two electrons occupying spatial orbitals $%
\left\vert a\right\rangle $ and $\left\vert b\right\rangle $ (which can be
understood as notation for one-dimensional irreducible representations of a
point group) and forming corresponding singlet and triplet states $^{1}B$
and $^{3}B$. The relevant wave functions in the coordinate representation
are given by:\
\begin{equation}
\begin{array}{c}
\Psi _{^{1}B}(x_{1},x_{2})=\frac{1}{2}\left( a(\mathbf{r}_{1})b(\mathbf{r}%
_{2})+b(\mathbf{r}_{1})a(\mathbf{r}_{2})\right) \left( \alpha (s_{1})\beta
(s_{2})-\beta (s_{1})\alpha (s_{2})\right) , \\ 
\Psi _{^{3}B}(x_{1},x_{2})=\frac{1}{2}\left( a(\mathbf{r}_{1})b(\mathbf{r}%
_{2})-b(\mathbf{r}_{1})a(\mathbf{r}_{2})\right) \left( \alpha (s_{1})\beta
(s_{2})+\beta (s_{1})\alpha (s_{2})\right) ,%
\end{array}
\label{MultipletFunctions}
\end{equation}%
both having the zero projection of the total spin. Following the definitions
of the one-electron density matrices eq. (\ref{DensityMatrices}) the states
eq. (\ref{MultipletFunctions}) immediately yield \emph{exactly} the same
one-electron density \emph{matrix}: 
\begin{equation}
\rho _{^{2S+1}B}^{(1)}(x,x^{\prime })=\frac{1}{2}\left( \alpha ^{\ast
}(s)\alpha (s^{\prime })+\beta ^{\ast }(s)\beta (s^{\prime })\right) \left(
a^{\ast }(\mathbf{r})a(\mathbf{r}^{\prime })+b^{\ast }(\mathbf{r})b(\mathbf{r%
}^{\prime })\right)  \label{MultipletOneElectronDensity}
\end{equation}%
irrespective to the total spin of these states. This result is well known
for decades and appears even in textbooks \cite{Zuelicke:book}. Obviously the
density eq.~(\ref{Density}) which is required by the DFT is as well the same
for the two spin states.

The \emph{exact} two electron density matrices calculated according to their
definition eq.~(\ref{DensityMatrices}) from the wave functions eq.~(\ref%
{MultipletFunctions}) \emph{are}, however, different: 
\begin{equation*}
\begin{array}{l}
\rho _{^{1,3}B}^{(2)}(x_{1}x_{2},x_{1}^{\prime }x_{2}^{\prime })= \\ 
\frac{1}{4}\left( \alpha ^{\ast }(s_{1})\beta ^{\ast }(s_{2})\mp \beta
^{\ast }(s_{1})\alpha ^{\ast }(s_{2})\right) \left( \alpha (s_{1}^{\prime
})\beta (s_{2}^{\prime })\mp \beta (s_{1}^{\prime })\alpha (s_{2}^{\prime
})\right) \times \\ 
\left( a^{\ast }(\mathbf{r}_{1})b^{\ast }(\mathbf{r}_{2})\pm b^{\ast }(%
\mathbf{r}_{1})a^{\ast }(\mathbf{r}_{2})\right) \left( a(\mathbf{r}%
_{1}^{\prime })b(\mathbf{r}_{2}^{\prime })\pm b(\mathbf{r}_{1}^{\prime })a(%
\mathbf{r}_{2}^{\prime })\right)%
\end{array}%
\end{equation*}%
with the upper sign corresponding to $S\!=\!0$ and the lower one to $S\!=\!1$%
. Comparing the above expression with the decomposition eq. (\ref{2Density})
one easily sees that only the \emph{cumulant} of the two-electron density
matrix can be responsible for the distinguishing of the two-electron density
matrices for the singlet and triplet states.

The\ "oversymmetry" of the density (and even of the first order density matrix) with respect to the total spin exemplified by eq. (\ref{MultipletOneElectronDensity}) is not accidental, but is a
consequence of a very general result (see Ref. \cite{Kaplan:07,Kaplan:07a} and
references therein). Even a higher symmetry can be proven \cite{Abarenkov:08}. In its modern formulation
(Theorem 1 of Ref. \cite{Kaplan:07}) it reads: 'The electron density of an arbitrary $N$-electron
system, characterized by the $N$-electron wave function corresponding to the
total spin $S$ and constructed on some orthonormal orbital set, does not
depend upon the total spin $S$ of the state and always preserves the same
form as it is for a single-determinant wave function'. The proof given in
Ref. \cite{Kaplan:07} relies not upon the spin properties themselves rather
on the manifestation of permutation symmetry of the exact wave function in
terms of the total spin. We address this issue later in Section \ref{DensityInUnitaryGroup}.

\subsection{Methods proposed to treat coinciding densities}

\subsubsection{Multiplet sum method}

The first attempt to get around this problem of coinciding densities in the
DFT context dates back to the work Ref. \cite{Ziegler:77}. The
analysis of problems performed there is precisely repeated in the above
two-electron two-orbital model. The prescription Ref. \cite%
{Ziegler:77} concerning the way out reads as follows: to evaluate
correctly the energy of the singlet and triplet states $^{1}B$ and $^{3}B$
in terms of the quantities which can be obtained with use of single
determinant wave functions. To do so one has to address the single
determinant function $\left\vert a\alpha b\beta \right\vert $ which is not a
pure spin state, but in fact is a linear combination of two above spin
states: 
\begin{equation}
\left\vert a\alpha b\beta \right\vert =\frac{1}{\sqrt{2}}\left( \left\vert
^{1}B,S_{z}=0\right\rangle +\left\vert ^{3}B,S_{z}=0\right\rangle \right)
\label{MSM-example}
\end{equation}%
Averaging the Hamiltonian over the linear combination eq. (\ref{MSM-example}%
) of the pure spin states immediately yields:%
\begin{equation}
\frac{1}{2}E(^{1}B)+\frac{1}{2}E(^{3}B)  \label{MSM-example-Energy}
\end{equation}%
The energy of the triplet state entering the above combination can be
independently extracted from another single determinant wave function: $%
\left\vert a\alpha b\alpha \right\vert $ corresponding to the component of
the triplet with the spin projection $+1$. Thus one can express the energy
of the (non-single-determinant) singlet state linearly combining the
averages of the Hamiltonian over the single determinant states of which,
however, one belongs to the spin projection $+1$. Obviously the above move
was only possible because the off-diagonal matrix element of the Hamiltonian
between the singlet and triplet contributions to the determinant of interest
vanishes due to the spin symmetry. The different expressions for $E(^{1}B)$
and$\ E(^{3}B)$ thus obtained are then treated as required \emph{distinct}
energy functionals to be used to calculate the energy respectively for the
singlet and triplet states possessing the same one-electron density. It is
instructive to check (and in this simple case it can be done by direct
evaluation) where the difference between the energy expressions comes from.
Inserting the one-electron density matrix eq. (\ref%
{MultipletOneElectronDensity}) which is the same for both spin states in the
definitions of the Hartree and exchange energies eqs. (\ref{HartreeEnergy}),
(\ref{ExchangePart}) yields for the both spin states equal Hartree and
exchange contributions:%
\begin{equation}
\begin{tabular}{ll}
Hartree & $\frac{1}{2}\left[ (aa|aa)+(aa|bb)+(bb|aa)+(bb|bb)\right] ;$ \\ 
exchange & $\frac{1}{2}\left[ (aa|aa)+(ab|ba)+(ba|ab)+(bb|bb)\right] .$%
\end{tabular}%
\end{equation}%
One can see that (i) the self interaction terms in the Hartree contribution
are precisely cancelled by the corresponding terms in the exchange part;
(ii) at the same time, obviously, there is no other source where the
difference between the spin state energies could come from except the
cumulant of the two-electron density matrix and thus the correlation energy
as defined by eq. (\ref{CorrelationPart}) is responsible for the difference
in the resulting expressions: 
\begin{equation}
\begin{array}{rr}
E(^{1}B) & =(aa|bb)+(ab|ba) \\ 
E(^{3}B) & =(aa|bb)-(ab|ba)%
\end{array}
\label{State-Energies}
\end{equation}%
On the other hand one may notice that the classification of the energy
contributions as exchange or correlation ones by eqs. (\ref{ExchangePart}), (%
\ref{CorrelationPart}) is in some way arbitrary as well. Indeed, for the
above model the energy of the triplet state with the spin projection $+1$ is
exactly the sum of the Hartree and exchange contributions since the latter
state is represented by a single determinant wave function for which the
cumulant precisely vanishes. However, the equal energy for the triplet state
with the zero spin projection breaks down differently: into the Hartree,
exchange, and correlation contributions, where the Hartree contribution is
the same as in the case of the spin projection $+1$, but only the sum of the
exchange and correlation contributions is the same for different values of $%
S_{z}$.

The above way leading to the energy expressions for different spin states is
not completely satisfactory (and not clearly generalizable): although
formally the results can be treated as functionals of the density the
difference of the two energy expressions is obtained by a kind of trick.
Referring to the triplet component with $S_{z}=+1$ in the derivation of the
multiplet energy looks out as an alien element (in fact the energy is
uniquely determined by the spatial multipliers in the wave functions eq. (%
\ref{MultipletFunctions}) -- without any reference to the spin components at
all). This strange element of the derivation appeared in order to compensate
somehow the element of the general theory which is missing in the DFT -- the
cumulant of the two-electron density matrix. Despite this criticism the
result of the derivation is very transparent: it reduces to deriving
according to McWeeny's notice in Ref. \cite{McWeenybook2nd} 'of a particular
type of energy expression---irrespective of the nature of the wavefunction',
namely one -- linear in the Coulomb and exchange two-electron integrals over
the involved orbitals.

Further development of this approach is based on the \emph{assumption} that
it is \emph{always} (or at least for unspecifically wide class of cases)
possible to express the energy of a pure spin multiplet\ state allowable for
a given number of electrons/orbitals as a linear combination: 
\begin{equation}
E(n\Gamma S)=\sum_{i}w_{i}^{n\Gamma S}E_{i}  \label{MultipletSumRule}
\end{equation}%
where $E_{i}$ are the diagonal matrix elements of the energy operator taken
with respect to all necessary Slater determinants (numbered by $i$). The
one-electron density (matrices) corresponding to these determinants are
different and the whole scheme becomes workable provided the set of the
coefficients (weights) $w_{i}^{\Gamma S}$ exists and they are uniquely
defined by the spin and symmetry quantum numbers $\Gamma $ and $S$ and other
quantum numbers $n$ serving to distinguish potentially existing states with
equal $\Gamma $ and $S$. Apparently only a restricted number of examples of
such functional forms is known. The reason is quite simple and the above
consideration allows to single out the range of cases where the derivation
analogous to that of eq. (\ref{MSM-example}) can be performed. It applies if
the total spin allows to completely distinguish the electronic states. In
this case the energy of the state of the highest available spin $S_{\max }$
can be expressed through a single determinant function (the cumulant is
vanishing) with the highest available projection of the total spin. Then
this result can be used to evaluate the energy of the state with $S_{\max
}-1 $, \textit{etc. }The recipe immediately fails as soon as multiple states
of the same total spin appear in the system. This is however the everyday
life, so in what follows we switch to considering further possibilities of
constructing the energy functionals useful in this situation.

\subsubsection{Restricted open shell KS (ROKS) method}

The situation with reproducing total spin dependence of the energy as it
appears in the DFT context is by no means unique: the same problem arises in
the Hartree-Fock-Roothaan (HFR) context since the latter lacks any adequate
representation of the cumulant of the two-electron density matrix as well.
Within the "extended" HFR context some ways out have been proposed.
Incidentally, the method of Ref. \cite{Ziegler:77} is precisely the
Slater multiplet sum method Ref. \cite{Slater:72} which migrated from the
HFR to the DFT context. Another option is the ROHF (restricted open shell
Hartree - Fock) method whose respective migration resulted in a range of the
ROKS (restricted open shell Kohn-Sham procedures \cite%
{Russo:94,Filatov:98}) being the DFT counterpart of the former.
Despite different appearance they have many common features (and we do not
address here the methods based on the statistical -- ensemble -- averaging).

The derivation of the ROHF (or equivalently \ 'old MC SCF ' -- see below)
bases on the general expression of the form: 
\begin{equation}
E(n\Gamma S)=\sum_{ij}C_{i}^{n\Gamma S}C_{j}^{n\Gamma S}H_{ij}
\label{RoothaanMethod}
\end{equation}%
where $C_{i}^{n\Gamma S}$ are the expansion coefficients of the
eigenfunction $\Psi _{n\Gamma S}(x_{1},x_{2},x_{3},\ldots x_{N})$ of the
many-electron hamiltonian over some appropriate basis states $\Phi
_{i}(x_{1},x_{2},x_{3},\ldots x_{N})$ (\textit{e.g.} the Slater
determinants, see, however, below). In this case no alien states of wrong
spin projection may appear. On the other hand the contribution of the
off-diagonal elements $H_{ij}$ to the energy may be nontrivial (in contrast
with eq. (\ref{MultipletSumRule})). The knowledge of the expansion
coefficients $C_{i}^{n\Gamma S}$ in general requires diagonalization of the
Hamiltonian matrix making the expansion coefficients and thus the energy
itself some sophisticated irrational function of the Hamiltonian matrix
elements including two-electron integrals. It was Roothaan \cite%
{Roothaan:60} who first noticed that certain states $\Psi _{n\Gamma S}$ of
atoms and linear molecules, even those requiring many-determinant
(multi-reference, multi-configurational) wave functions, yield energy
expressions which are linear with respect to two-electron integrals $\left(
ii|jj\right) $ and $\left( ij|ji\right) $ (respectively Coulomb and exchange
ones). It is only possible if the wave function expansion coefficients $%
C_{i}^{n\Gamma S}$ in eq. (\ref{RoothaanMethod}) can be determined on the
symmetry grounds \textit{i.e.} without nontrivial diagonalization. In this
case there is no need that the off-diagonal elements $H_{ij}$ which are
linear expressions in the two-electron integrals and thus give a linear
contribution to the energy functional disappear as required by eq. (\ref%
{MultipletSumRule}). Only the possibility to have the expansion coefficients 
$C_{i}^{n\Gamma S}$ independent on the specific values of the Hamiltonian
matrix elements is the true prerequisite for obtaining the expressions for
the energy of the required (linear) form. Nevertheless the number of cases
when the described procedure was possible in fact reduces to the $p^{n}$
states of atoms and $\pi ^{n}$ and $\delta ^{n}$ states of linear molecules.
Similarly the ROKS sheme proposed in Ref. \cite{Filatov:98} and
representing a migration of the Roothaan's reasoning to the DFT context
allowed to obtain the functional forms for the same set of states: $p^{n},$ $%
\pi ^{n},$ and $\delta ^{n}$. Thus the forecast of the year 1960 due to
Roothaan \cite{Roothaan:60}:\ 'It is a relatively simple matter to extend
the open-shell theory just presented in such a way that other important
classes of atomic states can be accommodated, as for instance, the $d^{N}$
configurations for the transition elements. We postpone such generalizations
for the present, and include whatever new treatments may be necessary with
the actual applications planned for the future' never became true and the $%
d^{N}$ states generally cannot be squeezed in the ROHF/ROKS scheme.

Under other angle of view, validity  of the Roothaan or  similar
schemes means that the cumulant of the two-electron density matrix can be in
some particular case recovered by symmetry based manipulations. In the cases
considered by Roothaan himself and recently used in the DFT context in Ref. 
\cite{Filatov:98} the possibility of obtaining closed expressions for the
energy functional in terms of two-electron integrals over orbitals involved
is stipulated by additional symmetry of the system (in the chemical context
it goes about additional symmetry group $G$ with irreducible representations 
$\Gamma $, where $G=SO(3)$ for an atom $G=SO(2)$\ for a linear molecule, and
may be some point group for other molecules) which allows to figure out the
expansion coefficients $C_{i}^{n\Gamma S}$. It is clear that for an
overwhelming majority of cases it is impossible to find any nontrivial symmetry group $G\neq C_{1}$ which predefines a restricted character of any Roothaan-like
treatment. It is thus our next purpose is to expore other possibilities of
designing energy functionals distinguishing the states of different spin
multiplicities in the DFT context.

\section{Spin and unitary symmetry of the electronic wave function\label%
{Unitary}}

As we mentioned previously any reference to the spin projections throughout
the derivation of the energy expressions for the two-electron two-orbital
model looks out as an alien element. The ultimate reason for that is that
the non-relativistic Hamiltonian does not depend on spin variables at all
and the energy itself as well as the differences in its form for different
spin states originates solely from the spatial multiplier of the
many-electronic wave function (spatial function). The idea to restrict the
entire consideration by those spatial functions persists almost from the
beginning of the quantum chemistry and is known as \ "spin-free quantum
chemistry" \cite{Matsen:64}. It can be given different formulations of which we
use one based on the use of the unitary group (see Ref. \cite{McWeenybook2nd}%
). We briefly remind its basic facts in the Appendix.

\subsection{Unitary symmetry of the spatial multiplier.}

The construct\ using the permutational symmetry of the spatial part of the
wave function had been used for developing the so called generalized
Hartree-Fock procedure \cite{Goddards} which had numerous descendants (see 
\emph{e.g.} \cite{Vojtik:72,Sullivan:72}). They basically performed the
task of presenting the energy in the HFR-like form: linear with respect to
Coulomb and exchange integrals over the orbitals involved with the
coefficients dependent on the permutational symmetry of the spatial part of
the wave function and thus on the total spin. The permutational symmetry,
however, addresses the many-electron wave functions in the coordinate
representation which is of restricted use in quantum chemistry. By contrast
the wave functions actually used are those in the representation of the
occupation numbers of the orbitals involved. For that reason it is more
practical to switch to labeling of the many-electron functions by
irreducible representations of the unitary group which are closely related
to those of the $S_{N}$ group. The corresponding construct is described in
the Appendix.

\subsection{Physical quantities in terms of unitary group.}

Going to the representation of the $U(M)$ group has that advantage that it
allows to easily write down the energy of many electron states. This is done
as follows: for each Young pattern $\Upsilon $ one can construct the set of
generators $\mathbf{E}_{ij}^{\Upsilon }$ ($ij=1\div M$) of the group $U(M)$
acting in the space of the irreducible representation $\Upsilon =\Upsilon
(M,N,S)$ whose matrix elements between the tableaux $\upsilon $ and $%
\upsilon ^{\prime }$ can be calculated irrespective to the physical nature
of the system described. The set of generators completely defines the action
of the group $U(M)$ in the irreducible subspace of its tensors of the rank $%
N $ with the permutational/spin symmetry stipulated by the Young pattern $%
\Upsilon $.

The diagonal generators $\mathbf{E}_{ii}^{\Upsilon }$ are diagonal in the
basis of Young tableaux and their matrix elements are equal to the
occupation number ($n_{i}=2,1,0$) of the $i$-th orbital in the Young tableau 
$\Upsilon \upsilon :$%
\begin{equation*}
\left\langle \Upsilon \upsilon \left\vert \mathbf{E}_{ii}^{\Upsilon
}\right\vert \Upsilon \upsilon ^{\prime }\right\rangle =\delta _{\upsilon
\upsilon ^{\prime }}\left\langle \mathbf{E}_{ii}^{\Upsilon }\right\rangle
_{\Upsilon \upsilon }=\delta _{\upsilon \upsilon ^{\prime }}n_{i}
\end{equation*}%
By contrast off-diagonal generators $\mathbf{E}_{ij}^{\Upsilon }$ (raising
ones if $i>j$ and lowering ones if $j>i$) have non-vanishing matrix elements $%
\left\langle \Upsilon \upsilon \left\vert \mathbf{E}_{ij}^{\Upsilon
}\right\vert \Upsilon \upsilon ^{\prime }\right\rangle $ if the tableau $%
\upsilon ^{\prime }$ contains at least one orbital symbol $j$ whereas the
tableau $\upsilon $ contains one less orbital symbol $j$ than $\upsilon
^{\prime }$ and one more orbital symbol $i$ than it. From this it follows
that the off-diagonal generators $\mathbf{E}_{ij}^{\Upsilon }(i\neq j)$ have
no non-vanishing diagonal matrix elements.

\subsubsection{One-electron density in the unitary group formalism}
\label{DensityInUnitaryGroup}

The generators $\mathbf{E}_{ij}^{\Upsilon }$ are \emph{by definition} the
components of the spatial one-electron density operator restricted to the
subspace of the states belonging to the $\Upsilon $ pattern (those having
transformation properties of the corresponding rank $N$ tensors with the
permutational symmetry stipulated by the Young pattern $\Upsilon $ or
equivalently having the total spin prescribed by this pattern):%
\begin{equation*}
\mathbf{E}_{ij}^{\Upsilon }=\dsum\limits_{\sigma }P^{\Upsilon }\mathsf{\psi }%
^{+}(i\sigma )\mathsf{\psi }(j\sigma )P^{\Upsilon },
\end{equation*}%
where $P^{\Upsilon }$ stands for the operator projecting $N$-electron wave
function to the subspace of the functions with the spatial part having the
permutational symmetry of the irreducible representation $\Upsilon $ of
either $S_{N}$ or $U(M)$ groups. It is remarkable to note that the Young
tableaux states have an important property similar to that of the Slater
determinants: the one-electron density matrices generated from such states
are diagonal in the basis of the orbitals involved in their construction.

With use of this construct one can easily check the validity of the Kaplan's
Theorem 1. Indeed, for whatever Young tableau $\Upsilon \upsilon $ the one
electron density pertinent to the corresponding $N$-electron state reads:%
\begin{eqnarray}
\rho _{\Upsilon \upsilon }(\mathbf{r,{r}}^{\prime }) &=&\sum\limits_{s}\rho
_{\Upsilon \upsilon }^{(1)}\mathbf{({r}}s\mathbf{,{r}}^{\prime }s\mathbf{)=}%
\sum\limits_{s}\left\langle \Upsilon \upsilon \left\vert \mathsf{\psi }^{+}(%
\mathbf{r}s\mathbf{)}\mathsf{\psi }\mathbf{({r}}^{\prime }s)\right\vert
\Upsilon \upsilon \right\rangle  \label{KaplanTheorem1} \\
&=&\dsum\limits_{\sigma }\sum\limits_{s}\sigma ^{\ast }(s)\sigma
(s)\dsum\limits_{ij}\varphi _{i}^{\ast }(\mathbf{{r})}\varphi _{j}\mathbf{({r%
}}^{\prime })\left\langle \Upsilon \upsilon \left\vert \mathsf{\psi }%
^{+}(i\sigma )\mathsf{\psi }(j\sigma )\right\vert \Upsilon \upsilon
\right\rangle =  \notag \\
&=&\dsum\limits_{ij}\varphi _{i}^{\ast }(\mathbf{{r})}\varphi _{j}\mathbf{({r%
}}^{\prime })\left\langle \Upsilon \upsilon \left\vert \mathbf{E}%
_{ij}^{\Upsilon }\right\vert \Upsilon \upsilon \right\rangle =  \notag \\
&=&\dsum\limits_{ij}\varphi _{i}^{\ast }(\mathbf{{r})}\varphi _{j}\mathbf{({r%
}}^{\prime })\delta _{ij}\left\langle \Upsilon \upsilon \left\vert \mathbf{E}%
_{ii}^{\Upsilon }\right\vert \Upsilon \upsilon \right\rangle  \notag \\
&=&\dsum\limits_{i}n_{i}\varphi _{i}^{\ast }(\mathbf{{r})}\varphi _{i}%
\mathbf{({r}}^{\prime })  \notag
\end{eqnarray}%
which in turn does not depend on the permutation symmetry labels $\Upsilon
\upsilon $, which is the only connection to the total spin. Thus even the
spatial density matrix (not only the density) is permutation/spin
independent as stated in Ref. \cite{Kaplan:07}.

\subsubsection{Energy in the unitary group formalism}

Further development is based on the possibility to express the blocks of the
Hamiltonian matrix pertaining to $N$ electrons in $M$ orbitals with total
spin $S$ through the generators $\mathbf{E}_{ij}^{\Upsilon }$,\ with $%
\Upsilon =\Upsilon (M,N,S)$. The required representation reads Ref. \cite%
{McWeenybook2nd}: 
\begin{eqnarray}
\mathbf{H} &\mathbf{=}&\bigoplus\limits_{\Upsilon }\mathbf{H}^{\Upsilon }
\label{ElectronHamiltonianUnitaryGroup} \\
\mathbf{H}^{\Upsilon } &=&\sum\limits_{ij}h_{ij}\mathbf{E}_{ij}^{\Upsilon }+%
\frac{1}{2}\sum\limits_{\substack{ ijkl}}(ij|kl)\left( \mathbf{E}%
_{ij}^{\Upsilon }\mathbf{E}_{kl}^{\Upsilon }-\delta _{jk}\mathbf{E}%
_{il}^{\Upsilon }\right)
\end{eqnarray}%
The matrix elements $h_{ij}$ are the sums of the respective matrix elements
of the kinetic energy $\hat{T}_{e}$ of electrons and of the external Coulomb
potential $\hat{V}_{ne}$; the quantities $(ij|kl)$ -- the two-electron
matrix elements of the Coulomb interactions.

For each of the states $\left\vert \Upsilon \upsilon \right\rangle $
represented by the Young tableau with the Young pattern $\Upsilon $ and the
filling $\upsilon $ (this information suffice to define the spatial part of
the $N$-electron wave function) the expectation value of the energy reads:%
\begin{equation}
E(\Upsilon \upsilon )=\sum\limits_{ij}h_{ij}\left\langle \mathbf{E}%
_{ij}^{\Upsilon }\right\rangle _{\Upsilon \upsilon }+\frac{1}{2}\sum\limits
_{\substack{ ijkl}}(ij|kl)\left\langle \left( \mathbf{E}_{ij}^{\Upsilon }%
\mathbf{E}_{kl}^{\Upsilon }-\delta _{jk}\mathbf{E}_{il}^{\Upsilon }\right)
\right\rangle _{\Upsilon \upsilon }  \label{EnergyThroughGenerators}
\end{equation}%
For the one-electron contribution to the energy one gets: 
\begin{equation*}
\sum\limits_{ij}h_{ij}\left\langle \mathbf{E}_{ij}^{\Upsilon }\right\rangle
_{\Upsilon \upsilon }=\sum\limits_{i}h_{ii}\left\langle \mathbf{E}%
_{ii}^{\Upsilon }\right\rangle _{\Upsilon \upsilon
}=\sum\limits_{i}h_{ii}n_{i}
\end{equation*}%
and the Coulomb interaction of electrons is expressed through the Coulomb
and exchange integrals with respect to the orbitals involved in the
construction of the states represented by the Young tableaux:%
\begin{equation}
\begin{array}{ll}
\mathrm{Hartree} & \frac{1}{2}\sum\limits_{ij}\left( ii|jj\right)
\left\langle \mathbf{E}_{ii}^{\Upsilon }\mathbf{E}_{jj}^{\Upsilon
}\right\rangle _{\Upsilon \upsilon }+ \\ 
\begin{array}{l}
\mathrm{exchange+} \\
\mathrm{correlation}%
\end{array}
& \frac{1}{2}\sum\limits_{i\neq j}\left( ij|ji\right) \left\langle \mathbf{E}%
_{ij}^{\Upsilon }\mathbf{E}_{ji}^{\Upsilon }-\mathbf{E}_{ii}^{\Upsilon
}\right\rangle _{\Upsilon \upsilon }-\sum\limits_{i}\left( ii|ii\right)
\left\langle \mathbf{E}_{ii}^{\Upsilon }\right\rangle _{\Upsilon \upsilon }%
\end{array}
\label{SelfEnergyThroughGenerators}
\end{equation}%
The Young tableau states $\Upsilon \upsilon $ are the eigenstates of the
diagonal generators $\mathbf{E}_{ii}^{\Upsilon }$. For that reason the
Hartree contribution to the energy can be rewritten:%
\begin{equation}
\begin{array}{ll}
\mathrm{Hartree} & \frac{1}{2}\sum\limits_{ij}\left( ii|jj\right)
\left\langle \mathbf{E}_{ii}^{\Upsilon }\right\rangle _{\Upsilon \upsilon
}\left\langle \mathbf{E}_{jj}^{\Upsilon }\right\rangle _{\Upsilon \upsilon }=%
\frac{1}{2}\sum\limits_{ij}\left( ii|jj\right) n_{i}n_{j}%
\end{array}%
\end{equation}%
in terms of the products of the one-electron densities. From this we see
that the Hartree part of the Coulomb energy is uniquely defined by the
occupation numbers of the spatial orbitals \textit{i.e.} only by the spatial
density in the representation of orbitals. We see that as in the other
representations the Hartree term is contaminated by the self-interaction of
electrons and that the principal effect for which the true exchange term is
responsible in the HFR context -- the avoiding of the self interaction -- is
guaranteed by the specific form of the coefficient at the integrals of the $%
\left( ii|ii\right) $ type which must be absorbed by the exchange
contribution to the energy. The averages of the off-diagonal generators'
products entering the expression eq. (\ref{SelfEnergyThroughGenerators}) are
not however uniquely defined either by the occupation numbers of the
orbitals in the tableau $\Upsilon \upsilon $ or by the total spin,
prescribed by the pattern $\Upsilon $. They depend also on the mutual
positions of the orbital symbols in the tableau. This is precisely the
result obtained many years ago in Ref. \cite{Poshusta:68} under the
name of the spin-free self consistent field theory.

>From the ROHF (old MCSCF) point of view the result eqs. (\ref%
{EnergyThroughGenerators}), (\ref{SelfEnergyThroughGenerators}) can be
considered as a recipe of obtaining the coupling coefficients $a_{ij}$ and $%
b_{ij}$ at the Coulomb and exchange integrals in the ROHF expressions for
the energy which incidentally acquire the $\Upsilon \upsilon $\ dependence:%
\begin{eqnarray}
a_{ij}^{\Upsilon \upsilon } &=&\left\langle \mathbf{E}_{ii}^{\Upsilon }%
\mathbf{E}_{jj}^{\Upsilon }-\delta _{ij}\mathbf{E}_{ii}^{\Upsilon
}\right\rangle _{\Upsilon \upsilon }  \label{RoothaanCoupling-a} \\
b_{ij}^{\Upsilon \upsilon } &=&\left\langle \mathbf{E}_{ij}^{\Upsilon }%
\mathbf{E}_{ji}^{\Upsilon }-\mathbf{E}_{ii}^{\Upsilon }\right\rangle
_{\Upsilon \upsilon }  \label{RoothaanCoupling-b}
\end{eqnarray}

Turning back to expressions eq . (\ref{RoothaanMethod}) one can say that
constructing the spatial Young tableaux states provide the expansion
coefficients $C_{i}^{\Upsilon \upsilon }$ for the respective linear
combinations of the $N$-electron Slater determinants, yielding the total
spin specified by the Young pattern. These coefficients are derived by
purely symmetry reasons and do not depend on the matrix elements of the
Hamiltonian thus satisfying the requirement of "universality". On the other
hand it is obvious that specifying the total spin only does not suffice to
specify the electronic state. The procedure implied by eq. (\ref%
{EnergyThroughGenerators}) provides for each allowable set of $M,N,S$ the
whole bunch of energy expressions labeled by the rows $\upsilon $ of the
irreducible representation $\Upsilon =\Upsilon (M,N,S)$.

\subsection{Multiplet sum method from the unitary perspective}

The first usage of the above formalism is to repeat the success of the MSM
in case of two electrons in two orbitals without addressing explicitly the
foreign component of the triplet state with $S_{z}=+1$. Indeed, the spatial
parts of the multiplet states in eq. (\ref{MultipletFunctions}) are
equivalently represented as the Young tableaux states:%
\begin{equation*}
\begin{array}{ll}
^{1}B & \left\vert 
\begin{tabular}{|l|l|}
\hline
a & b \\ \hline
\end{tabular}%
\;\right\rangle \\ 
&  \\[2pt] 
^{3}B & \left\vert 
\begin{tabular}{|l|}
\hline
a \\ \hline
b \\ \hline
\end{tabular}%
\;\right\rangle%
\end{array}%
\end{equation*}%
Two electrons in two orbitals form only one spatial function for the spin
triplet state, but in addition to one given above two more functions
compatible with the spin singlet state: 
\begin{equation*}
\left\vert 
\begin{tabular}{|l|l|}
\hline
a & a \\ \hline
\end{tabular}%
\;\right\rangle ,\left\vert 
\begin{tabular}{|l|l|}
\hline
b & b \\ \hline
\end{tabular}%
\;\right\rangle
\end{equation*}%
are available. Three spatial functions compatible with the spin singlet
state together form a basis of the three-dimensional irreducible
representation of the group $U(2)$ corresponding to the total spin 0. The
single spatial function compatible with the spin triplet state spans the
one-dimensional irreducible representation of the group $U(2)$. The Young
pattern label $\Upsilon $ here can be replaced by indicating the total spin
only. Then the generator $\mathbf{E}_{ab}^{S=1}=0$, but for $S=0$ one has: 
\begin{equation*}
\left\langle 
\begin{tabular}{|l|l|}
\hline
a & a \\ \hline
\end{tabular}%
\left\vert \mathbf{E}_{ab}^{S=0}\right\vert 
\begin{tabular}{|l|l|}
\hline
a & b \\ \hline
\end{tabular}%
\;\right\rangle =\sqrt{2}=\left\langle 
\begin{tabular}{|l|l|}
\hline
b & b \\ \hline
\end{tabular}%
\left\vert \mathbf{E}_{ab}^{S=0}\right\vert 
\begin{tabular}{|l|l|}
\hline
a & b \\ \hline
\end{tabular}%
\;\right\rangle
\end{equation*}%
These values suffice to perform the matrix multiplication of the generators $%
\mathbf{E}_{ab}^{S=0}\mathbf{E}_{ba}^{S=0}$ in the general expressions eq. (%
\ref{SelfEnergyThroughGenerators}), so that we obtain for the contribution
of the average interaction to the energy:%
\begin{equation*}
\begin{array}{rr}
E(^{1}B) & =(aa|bb)+(ab|ba) \\ 
E(^{3}B) & =(aa|bb)-(ab|ba)%
\end{array}%
\end{equation*}%
as it should be. We see that the archetypical result is reproduced within the
Young tableaux technique without addressing the component of the spin
multiplet with a foreign value of the spin projection. Also the
self-interaction contamination is removed automatically.

\subsection{DFT implications}

All above treatment was not in any way related to the DFT realm. The
possibility of establishing such a relation can be based on the recognition
of the fact that the symmetry (in particular the total spin) dependence must
be extraneously introduced into  DFT considerations \cite%
{Barth:79,Gunnarson:76} analogously to the treatment by Filatov and
Shaik Ref.~\cite{Filatov:98}. The unitary group formalism allows us to
conclude that for a given set of consistent values of $M,N,S$ one arrives to
the family of functionals labelled by the rows $\upsilon $ of the
irreducible representation $\Upsilon =\Upsilon (M,N,S)$ of the $U(M)$
group.\ The spin symmetry features of these functionals are condensed in the 
$\left\langle \mathbf{E}_{ij}^{\Upsilon }\mathbf{E}_{ji}^{\Upsilon }-\mathbf{%
E}_{ii}^{\Upsilon }\right\rangle _{\Upsilon \upsilon }$ (or $%
a_{ij}^{\Upsilon \upsilon },$ $b_{ij}^{\Upsilon \upsilon }$) coefficients
given above.

The energy matrix elements reflecting specific features of the system 
can be easily figured out. The coefficients $a_{ij}^{\Upsilon
\upsilon }$ for the Coulomb integrals $\left( ii|jj\right) $ which 
define the Hartree part of the Coulomb energy are known, but they are of no practical use in the DFT context, where the Hartree part of the interaction 
is determined directly from the electron density. Relatively problematic (in the DFT context) is to decide where the energy matrix element to be combined with
the coupling coefficients $\left\langle \mathbf{E}_{ij}^{\Upsilon }\mathbf{E}%
_{ji}^{\Upsilon }-\mathbf{E}_{ii}^{\Upsilon }\right\rangle _{\Upsilon
\upsilon }$ (exchange) and $-\left\langle \mathbf{E}_{ii}^{\Upsilon
}\right\rangle _{\Upsilon \upsilon }$ (self-interaction) has to come from.
This choice must be compatible with various theoretical settings. First of
all we notice that if a hybrid functional is used which contains some
fraction of the Hartree-Fock exchange the latter must be modified
accordingly so that the corresponding $\left( ij|ji\right) $ integrals over
the Kohn-Sham orbitals be included with the correct coefficients $%
\left\langle \mathbf{E}_{ij}^{\Upsilon }\mathbf{E}_{ji}^{\Upsilon }-\mathbf{E%
}_{ii}^{\Upsilon }\right\rangle _{\Upsilon \upsilon }$. The same applies to
the integrals $\left( ii|ii\right) $ which together with coefficients $%
-\left\langle \mathbf{E}_{ii}^{\Upsilon }\right\rangle _{\Upsilon \upsilon }$
will take care about some fraction of self-interaction.

Further concerns are related with the treatment of the nontrivial parts of
the exchange-correlation functionals within the $\Upsilon \upsilon $
numbering of the spin (permutation) states. This can be solved on the basis
of certain consistency requirements. Among possible consistency requirements
the most natural is that with the TDDFT. The TDDFT approximation is
equivalent to constructing the corresponding time evolution of the
many-electronic state in the basis of single electron excitations
(particle-hole pairs) above the KS single determinant wave function. 
Leaving aside the question of the area of applicability
of such an approach we notice that it requires an estimate of the
two-electron integrals coupling between different singly excited 
determinants. The interaction appears as second functional derivative 
of the energy with respect to density (first
functional derivative of the exchange-correlation functional). In the orbital
representation these derivatives acquire the necessary form of
two-electron integrals with the kernels determined by the form of the used
exchange-correlation functional. On the other hand the $\left( ij|ji\right) $
integrals appear in ROHF\ and in unitary group formalism for $\Upsilon
\upsilon $ states as a consequence of configuration interaction between
different Slater determinants implicitly entering in the expansion of the
Young tableau state $\Upsilon \upsilon $. Thus in order to ensure the
compatibility of the corresponding components of the theory: the couplings
between the elementary excitations in TD-DFT and between Slater determinants
in expansions of $\Upsilon \upsilon $ states must be the same. Thus they can
be expressed through the integral kernels of the interaction (\textit{e.g.}
according to \cite{Rosa:04}):%
\begin{eqnarray}
\left( ij|ji\right) ^{\mathrm{xc}} &=&\dint \dint \varphi _{i}^{\ast }(%
\mathbf{r})\varphi _{j}(\mathbf{r})f^{\mathrm{xc}}(\mathbf{r},\mathbf{r}%
^{\prime })\varphi _{j}^{\ast }(\mathbf{r}^{\prime })\varphi _{i}(\mathbf{r}%
^{\prime })d\mathbf{r}d\mathbf{r}^{\prime };  \notag \\
&&\mathrm{where} \\
f^{\mathrm{xc}}(\mathbf{r},\mathbf{r}^{\prime }) &=&\frac{\delta v^{\mathrm{%
xc}}(\mathbf{r})}{\delta \rho (\mathbf{r}^{\prime })}  \notag
\end{eqnarray}%
Namely these quantities must be inserted in the expressions for the
exchange-correlation energies to get these later consistent with the total
spin/permutation symmetry of the underlying many-electronic ground state.
This is also in agreement with the way of constructing the coupling
operators by Filatov and Shaik in their version of ROKS Ref. \cite%
{Filatov:98} and a similar procedure can be easily designed for the $%
\Upsilon \upsilon $ labelled states.

\subsection{Further examples}

As we mentioned many times the spin in general does not suffice to
distinguish many electronic states with the same one-electron density which
produces problems in describing corresponding states in the DFT. The only
example of the usage of the unitary group formalism given so far was however
the simplest case when the total spin labeling was sufficient. Below we
briefly exemplify the features one should expect in general case when there
exist Young tableaux differing by the positions of the orbital symbols in
these tableaux. In this case one can say that for given $M,N,S$ uniquely
defining the irreducible representation $\Upsilon $ of the group $U(M)$ and
for the row $\upsilon $ of the latter defined by a specific location of the
orbital symbols in the tableau a "Hartree-Fock-like" energy functional can
be written whose electron-electron interaction part is given by eq. (\ref%
{SelfEnergyThroughGenerators}). It can be optimized with respect to the
expansion coefficients of the involved orbitals over the AO's basis yielding
an effective Fockian matrix whose eigenvectors are precisely the orbitals
involved in the construction of the Young tableau state in the same manner
as it is in the ROHF/ROKS.

It is easy to check that the positions of the orbital indices in the
tableaux really matter. For example, for two Young tableaux states: 
\begin{equation}
\left\vert \Upsilon \upsilon \right\rangle =\left\vert 
\begin{tabular}{|l|l|}
\hline
a & b \\ \hline
c & d \\ \hline
\end{tabular}%
\;\right\rangle ;\ \left\vert \Upsilon \upsilon ^{\prime }\right\rangle
=\left\vert 
\begin{tabular}{|l|l|}
\hline
a & c \\ \hline
b & d \\ \hline
\end{tabular}%
\;\right\rangle  \label{FourElectrons}
\end{equation}%
both representing singlet states of four electrons in four orbitals with
equal one-electron density matrices, the contributions to the energy
functionals of the form eq (\ref{SelfEnergyThroughGenerators}), proportional
to the exchange integrals, respectively, are \cite{KaplanSymmetry}: 
\begin{equation}
\begin{array}{rr}
\left\vert \Upsilon \upsilon \right\rangle : & (ab|ba)+(cd|dc)-\frac{1}{2}%
\left[ (ac|ca)+(ad|da)+(bc|cb)+(bd|db)\right] \\ 
\left\vert \Upsilon \upsilon ^{\prime }\right\rangle : & -(ab|ba)-(cd|dc)+%
\frac{1}{2}\left[ (ac|ca)+(ad|da)+(bc|cb)+(bd|db)\right]%
\end{array}
\label{YoungTableauxEnergies}
\end{equation}%
where the Hartree and the self-interaction correcting terms are omitted for
brevity.

Remarkably enough neither of the expressions eq.~(\ref{YoungTableauxEnergies}%
) (combined with other necessary temrs) yields a lower energy \textit{a
priori}: which one is lower depends on the relations between the molecular
integrals involved. At this point one can return to the qualitative
interpretation of the Young tableaux with different positions of the orbital
symbols as of reflecting different "pairing schemes". Indeed, the states in
eq. (\ref{FourElectrons}) can be respectively treated (and this is in accord
with the energy expressions eq. (\ref{YoungTableauxEnergies})) as pairwisely
coupling electrons in the states $a$ and $b$ and $c$ and $d$ to the singlets
and triplets then coupling these intermediate states to the final singlet
states.

On the other hand one can easily conclude that for the above pair of Young
tableaux $\Upsilon \upsilon $ and $\Upsilon \upsilon^{\prime }$ for which $%
n_{i}=n_{j}=1$ and the difference between them is only the positions of the
orbital symbols $b$ and $c$ in the tableaux eq. (\ref{FourElectrons}) the
operators $\mathbf{E}_{ij}^{\Upsilon }\mathbf{E}_{ji}^{\Upsilon }$ entering
as multipliers of the $\left( ij|ji\right) $ exchange integrals in the exact
Hamiltonian yield also an off-diagonal matrix element of the Hamiltonian 
\cite{KaplanSymmetry}:%
\begin{equation}
\left\langle \Upsilon \upsilon \left\vert \mathbf{H}^{\Upsilon }\right\vert
\Upsilon \upsilon^{\prime }\right\rangle =-\frac{\sqrt{3}}{2}\left(
(ac|ca)-(ad|da)-(bc|cb)+(bd|db)\right) \neq 0,  \label{CI-Matrix-Element}
\end{equation}%
which shows that in this case a 2$\times $2 diagonalization is required for
obtaining the electronic energy exactly. Thus the energy functional becomes
a square root irrational function of the two-electron integrals rather than
a linear one. Additional symmetry relations may produce the energy
expression linear in the Coulomb and the exchange integrals (in the above
example it suffice that exchange integrals $(ac|ca),(ad|da)$ and $%
(bc|cb),(bd|db)$ are pair-wisely equal (which makes eq. (\ref%
{CI-Matrix-Element}) be zero) or alternatively that the exchange integrals
satisfy the equality:

\begin{equation}
(ab|ba)+(cd|dc)=\frac{1}{2}\left[ (ac|ca)+(ad|da)+(bc|cb)+(bd|db)\right]
\end{equation}%
which makes the diagonal matrix elements of the Hamiltonian be equal for the
states represented by eq. (\ref{FourElectrons}). Both symmetries yield
specific forms of the 2$\times $2 configuration interaction matrix and by
this allow the diagonalization to be feasible on the purely symmetry
grounds).

\section{State-specific exchange-correlation functionals for atomic $d$%
-shells\label{d-shell}}

The above notion of irrationality shows that even the unitary group
formalism does not solve the problem of constructing density functionals for
the specific correlated states. Although the unitary group formalism allows
to significantly contract the expansions of the states of the definite total
spin (in fact the $\Upsilon \upsilon $ labeled states become
single-configuration, albeit each of them is a combination of many Slater
determinants) the nonlinearity of the energy expression with respect to the
two-electron integrals hinders constructing the symmetry adapted functionals
along the lines suggested above. This problem manifests itself in the
description of the many-electronic states in the $d$-shells of transition
metal ions. Using the unitary goup formalism also in this case does not
allow to go further than the Roothaan \emph{old} MC SCF theory as described
in Ref. \cite{Roothaan:60}.\ There the spin/angular momentum dependent
coupling coefficients $a_{ij}$ and $b_{ij}$ had been introduced ultimately
to express the cumulant of the two-electron density matrix using symmetry
considerations. They are valid only if the multiplet states can be uniquely
obtained by applying operators projecting the Young tableau states to the
specific rows of the irreducible representations of the $SO(3)$ or $SO(2)$
groups (atoms and linear molecules, respectively). In these two cases
moderately simple expressions for the classifying operators (respectively, $%
L^{2}$ and $L_{z}$) in terms of the generators $\mathbf{E}_{ij}^{\Upsilon }$
can be written and used for constructing the required symmetry adapted
combinations of the Young tableau states. In the case of the atomic $p$%
-shells (and molecular $\pi $- and $\delta $-shells) the number of the $%
SO(3) $ ($SO(2)$) symmetry labels (different values of the orbital momentum $%
L$) produced by the projection of the Young tableau states to the definite $%
L^{2} $ states suffice to distinguish all different energies in these
shells. In the case of atomic $p^{n}$-states the symmetry $SO(3)$ reduces
also the number of independent two-electron integrals (including bothe the
Coulomb and exchange ones) to only two independent Slater-Condon parameters $%
F_{k}(pp);k=0,2$. This allowed the authors of Ref. \cite{Nagys,Nagy:05} to develop
state-specific functionals for the atomic $p^{n}$-states. It turns out,
however, that for the $d$-shells it does not suffice for a major part of the
atomic electronic terms of the transition metal ions \cite{Ballhausen:62}. Even
in free ions where the multiple terms having the same spin and orbital
momentum do exist and their energies cannot be expressed linearly through
the two-electron integrals. Despite the high-symmetry situation of a free
atom (ion) which reduces all the two-electron integrals to  a limited number (three) Slater-Condon parameters $F_{k};k=0,2,4$ in
the free ions the energies of the multiplets require 2$\times $%
2-diagonalization and thus their analytical expressions contain square roots
(for a handy reference see \cite{Ballhausen:62}). This moment is crucial -- it
is not possible to get rid out of the irrationality (square root) in the
expression for the energy\ by linearly combining the parameters of the
Hamiltonian.

The situation clearly becomes less favorable in lower symmetries or in
larger subshells (\textit{e.g.} partially filled $f$-shells) where the terms
of the same spin and symmetry span the subspaces of dimensionalities higher
than two. For example, in the octahedral environment the LS states of $d^{4}$%
- ($d^{6}$-) configuration span up to seven-dimensional subspaces of
many-electronic states \cite{Sviridov:book}. Clearly, at an
arbitrarily low symmetry the problem of linearly expressing the exact energy
of many-electronic terms through the Coulomb and exchange integrals cannot
be solved and obviously the energy of any of such multiple terms cannot be
expressed as a linear combination of Coulomb and exchange integrals. In what
follows below we restrict ourselves to the case of atomic $d$-shells and the
square root irrationalities in the state-specific expressions for the energy
trying to squeeze the simplest thinkable irrationality reflecting nontrivial
correlations in a kind of generalized density functional.

\subsection{The example of Fe$^{2+}$ ion}

We concentrate on the free Fe$^{2+}$ ($d^{6}$) ion which is an important
object in the studies of biologically active transition metal complexes and
following \cite{Ballhausen:62} provides a rich system of nontrivially
correlated multiple states in its $d$-shell. Namely this kind of behavior is
known to systematically evade from any DFT-based treatment. The energy
expressions of the states in a free Fe$^{2+}$ ($d^{6}$) ion are given in
Table 1
. They nontrivially depend on two
Slater-Condon parameters: $F_{2}$ and $F_{4}$. The ground state follows the
Hund's rule and for the Fe$^{2+}$ (d$^{6}$) ion it is the $^{5}D$ state.
According to the data published Ref.~\cite{NIST} the states $_{\pm }^{1}S$
and $_{\pm }^{1}D$ are not resolved from the spectra. Also the $^{2}F$ state
cannot be present in the spectrum of an even-electron system. Thus we
exclude three uppermost rows of the Table 1 
and
finally arrive to the set of data Table 2 
which can be used for analysis.

In order to get an impression of what can be (and should be) possibly
achieved in terms of describing the multiple states of the $d$-shells we
determine parameters $F_{2}$ and $F_{4}$ from experimental data. This
can be done in a number of ways. The semi-empirical approach is to assume
that $F_{2}$ and $F_{4}$\ are independent parameters. At the first stage we
neglect the correlation and take into consideration only the average
energies of the $_{\pm }^{(2S+1)}L$ states. The corresponding set of
energies is given in Table 3
. These
energies are linear in the parameters $F_{2}$ and $F_{4}$. Applying the
standard linear least squares procedure yields the experimental
"non-correlated" estimate of the parameters (in cm$^{-1}$):

\begin{equation}
\begin{array}{ccc}
F_{2}^{\mathrm{\exp }} & = & 1411.0, \\ 
F_{4}^{\exp } & = & 120.25, \\ 
F_{2}^{\mathrm{\exp }}/F_{4}^{\exp } & = & 11.734.%
\end{array}
\label{noncorrelated-experimental}
\end{equation}%
The quality of this result can be assessed by the value of mean square
deviation which is $686.80$ cm$^{-1}$ which must be compared with the range
of the energies described by the model being \textit{ca.} 45000 cm$^{-1}$.

Next step consists in estimating the manifestations of correlations in the
available data set. The most direct way to do that is to consider the square
root contributions to the energies of the multiple terms of the same spin
and symmetry. These come from the diagonalization of the symmetry adapted CI
matrices. Technically the correlations of that sort are responsible for the
splitting within the pairs of states of the same spin and symmetry which do
not have any counterpart in the DFT and describe the nontrivial
(non-dynamical) part of the correlation. We can see from the Table 4 
that the correlation splitting between the double
states is by approximately 10\% underestimated when calculated with use of
the non-correlated experimental estimates of the $F_{2}^{\exp }$ and $%
F_{4}^{\exp }$ parameters eq. (\ref{noncorrelated-experimental}). The
overall picture as coming from the non-correlated estimate can be
characterized by its mean square deviation $1228.0$ cm$^{-1}$. This fit
seems to be improvable by performing another (nonlinear) one for the entire
set of available excitation energy expressions and the corresponding
experimental values as given in Table 2
. The result of this new fit is, however, twofold. The resulting values (cm$%
^{-1}$) of the parameters $F_{2}^{\exp }$ and $F_{4}^{\exp }$ eq.~(\ref%
{correlated-experimental}): 
\begin{equation}
\begin{array}{ccc}
F_{2}^{\mathrm{\exp }} & = & 1468.92, \\ 
F_{4}^{\exp } & = & 113.30, \\ 
F_{2}^{\mathrm{\exp }}/F_{4}^{\exp } & = & 12.960,%
\end{array}
\label{correlated-experimental}
\end{equation}%
which can be qualified as "correlated" experimental ones, produce the mean
square deviation of $842.37$ cm$^{-1}$\ which manifests a significant
improvement as compared to analogous usage of the "non-correlated"
experimental values eq.~(\ref{noncorrelated-experimental}). Meanwhile,
although the overall picture is improved the description of the average
multiplet energies is deteriorated as compared to the "non-correlated"
parameters eq.~(\ref{noncorrelated-experimental}) so that the corresponding
mean square deviation somewhat increases to the value of $859.37$ cm$^{-1}$.

The above results deserve thorough attention. First of all we notice
following Refs. \cite{Nagy:05,Hamamoto:04} that the parameters $F_{2}$ and $%
F_{4}$ are by definition some functionals of radial density: 
\begin{equation}
\begin{array}{ccc}
F_{k} & = & \displaystyle\frac{e^{2}}{D_{k}}\displaystyle\int\limits_{0}^{%
\infty }\int\limits_{0}^{\infty }\displaystyle\frac{\left[ \min (r_{1},r_{2})%
\right] ^{k}}{\left[ \max (r_{1},r_{2})\right] ^{k+1}}%
R^{2}(r_{1})R^{2}(r_{2})r_{1}^{2}r_{2}^{2}dr_{1}dr_{2} \\ 
&  & D_{0}=1;D_{2}=49;D_{4}=441%
\end{array}
\label{Slater-Condon}
\end{equation}%
where $R^{2}(r)$ is the radial density distribution for the involved atomic $%
d$-shell. However, according to the Theorem 2 of Ref.~\cite{Kaplan:07} for
whatever spatial multiplet the one-electron density is spherically
symmetric. Thus the quantities $F_{k}$ are the functionals of one-electron
density which in the said case have only the radial dependence $r=\left\vert 
\mathbf{r}\right\vert $. For that reason the energies in Table 1 
can be also treated as functionals of the
one-electron density representing the averages of the electron-electron
interaction energy for each specific many-electron state in the $d$-shell.
When supplied by the relevant one-electron contributions (expression for the
kinetic energy and that for the electron-nuclear attraction) they become the
state specific energy functionals 
\begin{equation}
T[R^{2}(r)]+V_{ne}[R^{2}(r)]+\frac{n_{d}\left( n_{d}-1\right) }{2}%
A[R^{2}(r)]+XC_{nLS}[R^{2}(r)]
\end{equation}%
where the contribution proportional to $A[R^{2}(r)]$ is remarkably analogous
to the Hartree energy, however, free from the self-interaction and the $%
XC_{nLS}$ contributions are the state specific exchange-correlation
functionals. They can be treated according to the variational principle (in
some analogy with Ref. \cite{Hamamoto:04}) this is going to yield some
integrodifferential equations for the functions $R(r)$. This option will be
considered in details elsewhere. Here we notice that assuming the model
Slater orbital form for the functions $R(r)$ in the $d$-shell: 
\begin{equation*}
R(r)=\frac{(2\zeta )^{n+\frac{1}{2}}}{\sqrt{(2n)!}}r^{n-1}\exp (-\zeta r)
\end{equation*}%
allows one to evaluate the integrals in eq.~(\ref{Slater-Condon}) thus
leading to the linear dependence of the latter on the orbital exponent $%
\zeta $: 
\begin{equation}
\begin{array}{ccc}
F_{0}^{\mathrm{th}} & = & \displaystyle\frac{793}{3072}\zeta \\ 
F_{2}^{\mathrm{th}} & = & \displaystyle\frac{2093\cdot 5}{49\cdot 76800}\zeta
\\ 
F_{4}^{\mathrm{th}} & = & \displaystyle\frac{91\cdot 9}{441\cdot 9216}\zeta%
\end{array}
\label{Slater-Condon-through-zeta}
\end{equation}%
(here $n=3$). The ratio of the theoretical values 
\begin{equation}
F_{2}^{\mathrm{th}}/F_{4}^{\mathrm{th}}=\frac{2093\cdot 5}{49\cdot 76800}/%
\frac{91\cdot 9}{441\cdot 9216}=13.8
\end{equation}%
can be compared with one extracted from the correlated and non-correlated
experimental estimates of the parameters $F_{2}^{\exp }$ and $F_{4}^{\exp }$
eqs. (\ref{correlated-experimental}), (\ref{noncorrelated-experimental})
which indicates that for some reasons the correlated model for the energies
better agrees with the Slater model for the radial density distribution.
Since as we mentioned the density is spherically symmetric for whatever of
the states listed in the above Tables the only parameter characterizing the
density is the orbital exponent $\zeta $, provided the said multiplets are
constructed on the Slater radial orbitals. In view of the linear dependency
of $F_{k}^{\mathrm{th}}$ on $\zeta $ the state specific expressions for the
energies and energy differences in Tables 1 
and %
2 
become linear functions of $\zeta $ as
well.

The excitation energy expressions can be converted to the full scale density
functionals for the $d$-shell if one complements the above electron
interaction energies by the one-electron terms for the $d$-shell with six
electrons in it. The one-electron terms are (i) the kinetic energy per
electron:%
\begin{equation}
\frac{\zeta ^{2}}{2};
\end{equation}%
(ii) the potential energy of attraction to the nucleus per
electron where the $3$ in the denominator stands for the principal quantum
number of the $d$-shell under consideration: 
\begin{equation}
-\frac{Z}{3}\zeta
\end{equation}%
The average electron-electron interaction value common for all
electronic terms is proportional to the Racah $A$ parameter whose expression
in terms of $F_{0}^{\mathrm{th}}$ and $F_{4}^{\mathrm{th}}$ is given in the
footnote to Table 1
. For the iron(II) ion we can set $Z=8$ and the number of $d$-electrons $n_{d}=6$
to take care about the core screening, then the expression for the energy
becomes:%
\begin{equation}
3\zeta ^{2}-16\zeta +15\frac{143}{576}\zeta .
\end{equation}%
The value of $\zeta $ comes then as one providing the minimum to the above
functional, so that: 
\begin{equation}
\zeta =\frac{8}{3}-\frac{5}{2}\frac{143}{576}\approx 2.0460...
\end{equation}%
in a remarkable correspondence with the Slater rules yielding for this
setting the value of $2.08$ simply by ascribing the screening increment of $%
0.35$ to each electron (except one) in the $d$-shell. The screening
increment coming form the formula for $A$ amounts $0.387$.

Including further contributions for the electron-electron interaction energy
which are now state specific yields for the ground state: 
\begin{equation}
\zeta \approx 2.0621
\end{equation}%
On the other hand taking one of the higher excited states $^{1}I$ whose
energy is about $30000$ cm$^{-1}$ above the ground state gives: 
\begin{equation}
\zeta \approx 2.0533
\end{equation}%
From these estimates one can derive the following conclusion: The orbital
exponent and thus the radial density is very weakly sensitive to whatever
correlations. This finding is in agreement both with the accepted concept of
correlation which attributes it exclusively to the cumulant of the
two-electron density matrix so that there is no need to reload its
manifestations on the density as well as with numerous demonstrations of 
\emph{no relation} between the one-electron density and the correlations
known in the literature (see \textit{e.g.} Ref.~\cite{Eschrig:book}).

Further analysis can be based on the observation that inserting the
theoretical definitions for the $F_{2}^{\mathrm{th}}$ and $F_{4}^{\mathrm{th}%
}$ parameters eq. (\ref{Slater-Condon}) into expressions for the excitation
energies result in linear models for these energies with the single fitting
parameter $\zeta $. Two such models can be constructed: the non-correlated
which uses only the average energies of multiple states with equal $L$ and $%
S $ and the correlated one which covers all ten available excitation
energies. Fitting the excitation energies to the non-correlated model yields
the value of $2.4823$ for $\zeta $. The quality of fitting with only one
parameter is certainly somewhat worse than that using two independent
parameters $F_{2}^{\mathrm{\exp }}$ and $F_{4}^{\mathrm{\exp }}$ and the
mean square deviation becomes $1058.0$ cm$^{-1}$ for the set of average
energies of the multiplets (non-correlated fit). The value of $\zeta $ which
comes from the linear fitting procedure with the correlated energy
expressions is$\ 2.4604$ and the mean square deviation is $1013.3$ cm$^{-1}$%
. We see that also in this case the correlations only marginally affect the
one-electron density distribution and that despite some deterioration of the
precision as compared with the two-parameter models the overall quality of
the fit is surprisingly good.

\subsection{Summary}

Let us summarize the findings of this Section. We managed to obtain
simple expressions for the energies of the nontrivially correlated ionic
states (these expressions include non-dynamic correlation through the square
root terms) with definite values of $L$ and $S$ as functions of a single
parameter $\zeta $ -- the Slater orbital exponent for the $d$-shell. In the
context of the accepted model it is the only quantity characterizing the
density in the $d$-shell. In a sense there is one-to-one correspondence
between the electron density of the $d$-shell and $\zeta $ thus the
expressions for the energy can be considered as state specific energy
functionals of the form:
\begin{equation}
n_{d}\frac{\zeta ^{2}}{2}-n_{d}\frac{Z}{n}\zeta +\frac{n_{d}\left(
n_{d}-1\right) }{2}\frac{143}{576}\zeta +XC_{nLS}(n_{d},\zeta )
\label{State-Specific-XC}
\end{equation}%
written in terms of the orbital exponent $\zeta $ uniquely related to the
density within the model used and where $XC_{nLS}(n_{d},\zeta )$ stands for
state dependent exchange-correlation terms as obtained by inserting the
expressions for the Slater-Condon parameters $F_{k}^{\mathrm{th}}$ eq. (\ref%
{Slater-Condon-through-zeta}) in the expressions given in Table 1 
or analogous expressions for other $d$-shell
fillings Ref.~\cite{Ballhausen:62}.

As one can see our estimates of the characteristic quantity $\zeta $ yield
the values which fall into two classes depending on the type of the
estimate: those coming from the variational estimate for the total energy of
each respective state give the values close to $\zeta =2.08$ coming from the
Slater rules. The estimates based on fitting the excitation energies to $%
\zeta $ yield much larger value (much less diffuse $d$-shell) about $2.5$
with with extremely weak influence of electron correlation on the estimates
of either of these types. These numerical results must be compared with
other (empirical) values of the orbital exponents. These, however,
demonstrate a wide range of values. \textit{E.g.} Ref. \cite%
{Clementi:63} report the value $\zeta =3.7266$; \ Ref. \cite{Burns:64}
suggests $\zeta =3.152$; \ Ref. \cite{Gouterman:66} gives $\zeta =2.722$;
and \ Ref. \cite{Bohm:81} provide $\zeta =3.15$ basically repeating the
value of Ref. \cite{Burns:64}. This indicates that either the correlated or
non-correlated estimates, coming from the excitation energies only, fall in
the range defined by the Slater rules and other semi-empirical estimates.
Although, the deviations between the density parameter estimates coming from
different types of procedures also are expectable 
(we remind the existence
of distinct thermochemical and spectral semi-empirical parameterizations) the
true source of observed deviations is of certain interest.

\section{Range-separated treatment of electronic \\ Coulomb
interaction in atomic $d$-shells \label{Short-Long}}

Based on the idea that the short-range behavior of the e-e interactions can be efficiently transferred from the homogeneous e-gas to arbitrary many-electron systems, while the long-range e-e interactions being much more system specific (less transferable),  Savin and Stoll suggested a generalization of the Kohn-Sham theory by splitting explicitly the short- and long-range e-e interactions~\cite{Stoll:85,Savin:96,Savin:96a,Leininger:97}.
The non-transferable long-range interactions can be assimilated to a wave function treatment, just like the kinetic energy in the conventional Kohn-Sham model, which results in a replacement of the non-interacting Kohn-Sham reference system by a "long-range-interacting" one. While in conventional KS theory the 
effective KS Hamiltonian has an exact single-determinant solution, the generalized, range-separated variant includes a certain amount of explicit e-e interaction and the corresponding effective Schrödinger equation has to be solved in a multi-determinant form~\cite{Toulouse:04a}. However, due to the nonsingular nature of the lr Coulomb operator, the solution can be converged considerably faster in both the one-electron and many-electron basis. 

Recent works on the range-separated hybrid methods were mostly based on the 
the separation of the Coulomb potential into short- and long-range parts was
performed according to:%
\begin{eqnarray}
\frac{1}{r_{12}} &=&\left( \frac{1}{r_{12}}\right) _{s}+\left( \frac{1}{%
r_{12}}\right) _{l}  \notag \\
\left( \frac{1}{r_{12}}\right) _{s} &=&\frac{\func{erf}(\mu r_{12})}{r_{12}}
\label{Short-Long-Erf} \\
\left( \frac{1}{r_{12}}\right) _{l} &=&\frac{\func{erfc}(\mu r_{12})}{r_{12}}
\notag \\
1 &=&\func{erf}(x)+\func{erfc}(x)  \notag
\end{eqnarray}%
The treatment of the long-range exchange has been done in the Hartree-Fock framework, while the correlation could be treated by MP2~\cite{Angyan:05} or CCSD(T)~\cite{Goll:05} level, leading to a successful description of London dispersion forces in vdW complexes~\cite{Gerber:07} or by MCSCF level~\cite{Fromager:07} to treat typical non-dynamic correlation problems, like the case of the H$_2$ dissociation. A simpler model, where long-and short-range correlations are both handled by density functional approximations (RSHX - exchange-only range separated hybrid~\cite{Gerber:05a}, like LC-$\omega$PBE of Scuseria~\cite{Vydrov:06d}) has been recently shown to be quite successful in predicting magnetic coupling constants in transition metal systems~\cite{Rivero:08}.

In the following, we examine the behavior of the range-separated approach on the simple Fe(II) ion model system.

\subsection{Range separated hybrid approach
\label{Short-Long-Gen}}

In order to make easier the evaluation of analytical integrals and obtain the $F_{k}^{\mathrm{th}}$ parameters, we decided to employ the "Yukawa"-like separation as proposed in Ref. \cite%
{Savin:95}:%
\begin{eqnarray}
\left( \frac{1}{r_{12}}\right) _{s} &=&\frac{\exp (-\beta r_{12})}{r_{12}},
\label{Short-Long-Yukawa} \\
\left( \frac{1}{r_{12}}\right) _{l} &=&\frac{1-\exp (-\beta r_{12})}{r_{12}}.
\notag
\end{eqnarray}%
The value of $\beta \rightarrow 0$ corresponds to the absence of the
long-range part. By contrast $\beta \rightarrow \infty $ corresponds to the
evanescence of the short-range part. The reach of the short-range interactions is roughly inversely proportional to the value of $\beta$ measured in inverse bohr units.

The initial assumption is  that only
the long-range part of the Coulomb interaction contributes to the
non-dynamical correlations in the $d$-shells 
so that only the matrix elements of $\displaystyle\left( \frac{1}{r_{12}}%
\right) _{l}$ must be taken into account when the CI matrices describe the
nontrivial correlation in the $d$-shells. In order to check this assumption
we have performed the following. With use of analytical results of Refs. \cite%
{Steinborn:75a,Steinborn:75b} to get for the Yukawa potential the following
expansion: 
\begin{eqnarray}
\left\vert \mathbf{r}_{1}-\mathbf{r}_{2}\right\vert ^{-1}\exp (-\beta
\left\vert \mathbf{r}_{1}-\mathbf{r}_{2}\right\vert ) &=&4\pi
\sum_{l=0}^{\infty }\sum_{m=-l}^{l}\left( r_{<}r_{>}\right) ^{-\frac{1}{2}%
}\\
&&I_{l+\frac{1}{2}}\left( \beta r_{<}\right) K_{l+\frac{1}{2}}\left( \beta
r_{>}\right) 
Y_{l}^{-m}\left( \frac{\mathbf{r}_{<}}{r_{<}}\right) Y_{l}^{m}\left( \frac{%
\mathbf{r}_{>}}{r_{>}}\right)  \notag
\end{eqnarray}%
where $I_{l+\frac{1}{2}}$ and $K_{l+\frac{1}{2}}$ are the modified Bessel
functions of the half-integer index, $r_{<}=\min (r_{1},r_{2})$, $r_{>}=\max
(r_{1},r_{2})$, and the vectors $\mathbf{r}_{<}$ and $\mathbf{r}_{>}$\ are
assigned correspondingly. The above expression must be inserted in the
definition of the matrix elements of the electron-electron interaction (see 
\textit{e.g.} Ref. \cite{Ballhausen:62}) which due to the spherical symmetry of
the Yukawa potential allows us to express these latter in terms of the short
range analogs of the Slater-Condon parameters. For the $3d$ Slater orbitals
with the orbital exponent $\zeta $ the estimates for the short range $F_{2}^{%
\mathrm{(s)}}$ and $F_{4}^{\mathrm{(s)}}$ and long-range $F_{2}^{\mathrm{(l)}%
}$ and $F_{4}^{\mathrm{(l)}}$\ contributions to the $F_{k}^{\mathrm{th}}$
parameters eq. (\ref{Slater-Condon-through-zeta}) are:

\begin{eqnarray}
F_{2}^{\mathrm{(s)}} &=&F_{2}f^{\mathrm{(s)}};F_{2}^{\mathrm{(l)}}=F_{2}f^{%
\mathrm{(l)}};f^{\mathrm{(l)}}+f^{\mathrm{(s)}}=1  \notag \\
F_{4}^{\mathrm{(s)}} &=&F_{4}g^{\mathrm{(s)}};F_{4}^{\mathrm{(l)}}=F_{4}g^{%
\mathrm{(l)}};g^{\mathrm{(l)}}+g^{\mathrm{(s)}}=1  \notag \\
f^{\mathrm{(l)}} &=&1-\frac{4\zeta ^{2}}{6279(\beta +2\zeta )^{12}}%
(1575\beta ^{10}+37800\beta ^{9}\zeta +413420\beta ^{8}\zeta ^{2}+  \notag \\
&&2714880\beta ^{7}\zeta ^{3}+11850720\beta ^{6}\zeta ^{4}+35848960\beta
^{5}\zeta ^{5}+75603840\beta ^{4}\zeta ^{6}+  \notag \\
&&107827200\beta ^{3}\zeta ^{7}+94591744\beta ^{2}\zeta ^{8}+38578176\beta
\zeta ^{9}+6429696\zeta ^{10}) \\
g^{\mathrm{(l)}} &=&1-\frac{4\zeta ^{2}}{91(\beta +2\zeta )^{12}}(63\beta
^{10}+1512\beta ^{9}\zeta +16380\beta ^{8}\zeta ^{2}+104832\beta ^{7}\zeta
^{3}+  \notag \\
&&433888\beta ^{6}\zeta ^{4}+1188096\beta ^{5}\zeta ^{5}+2101632\beta
^{4}\zeta ^{6}+2263040\beta ^{3}\zeta ^{7}+  \notag \\
&&1487616\beta ^{2}\zeta ^{8}+559104\beta \zeta ^{9}+93184\zeta ^{10}) 
\notag
\end{eqnarray}%
Introducing the new variable%
\begin{equation}
t=\frac{\beta }{\beta +2\zeta }
\end{equation}%
we get somewhat simpler expressions for the long-range scaling coefficients

\begin{eqnarray}
g^{\mathrm{(l)}} &=&-\frac{256t^{12}}{91}+\frac{2304t^{11}}{91}-\frac{%
8832t^{10}}{91}+\frac{18432t^{9}}{91}-240t^{8}  \notag \\
&&+144t^{7}-16t^{6}-\frac{144t^{5}}{7}-\frac{6t^{4}}{7}+\frac{30t^{3}}{7}+%
\frac{15t^{2}}{7}  \notag \\
f^{\mathrm{(l)}} &=&-\frac{6400t^{12}}{6279}+\frac{75520t^{11}}{6279}-\frac{%
134528t^{10}}{2093}+\frac{1285120t^{9}}{6279} \\
&&-\frac{384880t^{8}}{897}+\frac{14160t^{7}}{23}-\frac{23600t^{6}}{39}+\frac{%
2451808t^{5}}{6279}  \notag \\
&&-\frac{23150t^{4}}{161}+\frac{6910t^{3}}{483}+\frac{3455t^{2}}{483}  \notag
\end{eqnarray}%
Numerical optimization of the sum of square deviations, where the
theoretical values are obtained under the condition that the parameters $%
F_{2}$ and $F_{4}$ under the square roots are respectively replaced by the
long range contributions $F_{2}^{\mathrm{(l)}}$ and $F_{4}^{\mathrm{(l)}}$,
with respect to $\zeta $ and $t$ results in the values:%
\begin{eqnarray}
\zeta &=&2.46425  \notag \\
t &=&0.725436  \label{Long-Range-Result} \\
\beta &=&13.0218  \notag
\end{eqnarray}%
The  range separation parameter $\beta $\ is obtained by
inverting the definition of $t$. These values correspond to the following
 scaling parameters:%
\begin{eqnarray}
f^{\mathrm{(l)}} &=&0.965879 \\
g^{\mathrm{(l)}} &=&0.912733  \notag
\end{eqnarray}
The precision of this estimate can be characterized as previously by the mean
square deviation which amounts to $996$ cm$^{-1}$. Taking into account that the short-range e-e potential corresponding to $\beta=13.02$ falls down to a negligibly small value, say 0.001,  for $r=0.5$ bohr, it can be concluded that electron repulsion at shorter than 0.5 bohr direct space distance has an insignificant effect on the multiplet structure.

By contrast, if the theoretical values are obtained under the condition that the parameters $F_{2}$ and $F_{4}$ under the square roots are respectively replaced by the short range contributions $F_{2}^{\mathrm{(s)}}$ and $F_{4}^{\mathrm{(s)}}$, the optimization of the sum of square deviations with respect to $\zeta $ and $t$ results in the values:%
\begin{eqnarray}
\zeta &=&2.4653  \notag \\
t &=&0.0482271  \label{Short-Range-Result} \\
\beta &=&0.249838  \notag
\end{eqnarray}%
These values correspond to the magnitudes of the scaling parameters:%
\begin{eqnarray}
f^{\mathrm{(s)}} &=&0.982441 \\
g^{\mathrm{(s)}} &=&0.994545  \notag
\end{eqnarray}%
Incidentally, the precision of the procedure singling the
short-range part characterized as previously by the mean square deviation
yields  the value $985$ cm$^{-1}$, quite similar to the long-range estimate.
The reach of the "short-range" interactions, measured by analogous criteria as before (falling off the short-range Coulomb potential below 0.001) is about $r=16$ bohr, which englobes practically the full range for the significant densities of the $d$-electrons. It means that the long-range "tail" of the electron-electron interactions is essential to recover the correct muliplet structure. Furthermore, one can see that the renormalization of the $F$ functions is in the order of 1\%, confirming that the use of the optimal
$\beta$ implies the involvement of practically the full range of interactions   (cf.\ previous Section).

A further lesson drawn from this simple model study is that the
short range/long range separation of the Coulomb potential is
not sensitive to the correlations as well: the characteristic
parameter of the density distribution $\zeta $ in all cases equals to $2.46$
with variations in the third digit after the decimal point. Thus the
short range/long range separation does
not lift thus the strong contradiction between the estimates of the orbital
exponent by the Slater rules or variationally from the state specific
functionals eq. (\ref{State-Specific-XC}) and those from linear fit for the
excitation energies. Thunkable way out looks out twofold: First, the Slater rules
can be thought to overestimate the
screening (for the $d$-shell the screening by the inner shells is treated to
be complete, which yields the value $8$ for the effective charge) thus
leading to the values of $\zeta $\ too small as compared to those extracted
from the fitting of the experimental data on excitation energies. Second,
one can think
that the value of parameter $F_{0}$ is for some reason much stronger
renormalized as compared to its theoretical value eq. (\ref%
{Slater-Condon-through-zeta}) than those of the parameters $F_{2}$ and $%
F_{4} $. If we apply long-short range separation and calculate $F_{0}$
at the values of $\beta $ and $\zeta $ eq. (\ref{Short-Range-Result})
extracted from the fitting of excitation energies with the short-range parts
$F_{2}^{\mathrm{(s)}}$ and $F_{4}^{\mathrm{(s)}}$ under the square roots the
fraction of the short
range part in $F_{0}$ amounts $h^{\mathrm{(s)}}=0.688608$ of the latter. Now if we assume
that the short range part for some reason renormalizes to zero and thus
only the long range part of the e-e potetial contributes to the real value of $F_{0}$ then the variational
estimate of the orbital exponent reads:%
\begin{equation}
\zeta =\frac{8}{3}-\frac{5}{2}\cdot \frac{143}{576}\cdot (1-h^{\mathrm{(s)}%
})\approx 2.4743
\end{equation}%
which shows some reasonable consistency with the values extracted from analysis of
experimental spactra. Of course, this may well be a pure coincidence, but possible
consequences of the above hypothesis on the way of renormalization of the Slater-Condon parameters will be
considered elsewhere.

\section{Conclusion}

In the present paper we discussed a few possible ways of avoiding the
deadlocks of the pragmatic methods of molecular electronic structure theory
based on the DFT, which appear due to the non-sensitivity of the basic
quantity of the DFT -- the one-electron density -- to the differences in the spin (permutational) or/and spatial symmetry of the underlying many-electronic
states. This non-sensitivity is reflected by two theorems (recent
Theorems 1 and 2 of Ref. \cite{Kaplan:07}) which formalize two basically
known facts that (i)~the one-electron density does not depend on the total
spin of the many-electron state, and, that (ii)~the one-electron density in
a many-electronic state, which transforms according to any irreducible
representation of the group acting on the spatial coordinates of electrons
($SO(3),$ $SO(2)$, or their point subgroup), transforms according the fully
symmetric irreducible representation of the corresponding group.

These theorems imply that necessarily the information concerning
the symmetry of the respective many-electronic states at hand is to be
introduced into any DFT-based treatment extraneously. When it goes about the total spin (or equivalently about the permutational symmetry) of a
many-electron state, we suggest to use state-specific functionals labeled
by the Young tableaux $\Upsilon\upsilon $ (the rows of the irreducible representations of the  unitary group  $U(N)$) and to develop a procedure analogous to ROKS for each of them. In
the particular case of multiple states sharing the same $L$ and $S$
in the $d$-shells of transition metal ions we suggest state-specific
correlated functionals of the density and their model based on the
assumption of a Slater orbital
form of the radial density distribution. This procedure reduces the functional-type density dependence to function-type dependence on the orbital exponent. With the use of these expressions the excitation energies of the
many-electron states of the Fe$^{2+}$ ion are reproduced with remarkable accuracy. The
variational treatment of the proposed functionals reproduces with similar
precision the values of the orbital exponent of the Fe$^{2+}$ ion prescribed
by the Slater rules. Nevertheless, the estimates of the orbital exponent
coming from the variational principle and from the fit of the excitation
energies differ significantly although they fall in the range provided by
different semi-empirical estimates. Some ideas related to conciliation of
these two groups of estimates have been derived from analysis of the short
range/long range separation of the electron-electron interaction potential.

\section*{Acknowledgments}

The authors are thankful to Profs. I.V. Abarenkov, I. Mayer, G.M.
Zhidomirov, I.G. Kaplan, A. Savin, J.-P. Malrieu, \'{A}. Nagy for valuable
discussions. ALT expresses thanks to Ms O.A. Tchougr\'{e}eva for her
assistance. This work has been performed with financial support for ALT
coming from the RFBR grants Nos 07-03-01128, 08-03-00414.

\section*{Appendix. Permutation symmetry of the spatial function}

Since the Hamiltonian does not depend on spin variables one may wonder why
the total spin at all affects the energy. The answer lays in the symmetry of
many-electron wave functions with respect to permutations of\ coordinates $%
x_{i}$ of all $N$ electrons of the system (group $S_{N}$). The correct wave
function must be antisymmetric with respect to them (Pauli principle for
fermions). This simple statement applies when the complete electronic
coordinates $x_{i}=\left( \mathbf{r}_{i},s_{i}\right) ;i=1\div N$ are taken
as arguments of the wave function. Due to the fact that the nonrelativistic
electronic Hamiltonian does not depend on the spin projections $s_{i}$ the
wave function of electrons can be represented as a product of the spatial
and spin parts dependent respectively on the spatial ($\mathbf{r}_{i}$) and
spin ($s_{i}$) coordinates only (see \textit{e.g.} eq. (\ref%
{MultipletFunctions})) with the antisymmetry requirement applicable to the
entire products. In order to calculate the energy it is enough to know only
the spatial part (multiplier) of the $N$-electronic wave function. As one
can see in eq. (\ref{MultipletFunctions}) the \emph{spatial} parts of the
triplet and the singlet are respectively antisymmetric and symmetric with
respect to permutations of the spatial coordinates $\mathbf{r}_{1}$ and $%
\mathbf{r}_{2}$ and namely this difference is the only real source of the
differences in the energy.

In contrast with the simple permutation symmetry properties of the complete
wave functions those of the spatial multipliers are in general case somewhat
more involved. The permutation properties are conveniently described in
terms of the Young patterns and Young tableaux. Generally the Young patterns
are shapes formed by $N$ boxes arranged in rows of non-increasing length: 
\begin{equation*}
\Upsilon =%
\begin{tabular}{|l|l|l|l|l|l|l|}
\hline
&  &  &  &  &  &  \\ \hline
&  &  &   \\ \cline{1-4}
&  &  &   \\ \cline{1-4}
&  &  \\ \cline{1-3}
&  &  \\ \cline{1-3}
\end{tabular}%
\end{equation*}%
These shapes label the irreducible representations of the group $S_{N}$. The
Young pattern corresponding to the totally antisymmetric wave function
contains only one column of the height $N$. The fact that there is only one
possibility to fill this column by electron labels from $1$ to $N$
corresponds to the one-dimensionality of the antisymmetric representation of
the $S_{N}$ group. Separation of the antisymmetric function into spatial and
spin parts predefines their respective permutation properties: they must
belong to the adjoint representations of the $S_{N}$ group, since the
product of two functions belonging to adjoint representations yields the
required antisymmetric function. The Young patterns corresponding to adjoint
representations of $S_{N}$ are connected by 180$^{\circ }$ rotation around
the bissectriss of their common upper left corner. Since the Young patterns
which can be used for constructing the spin functions may contain no more
than two rows, those usable for constructing the electronic spatial
functions respectively cannot contain more than two columns: 
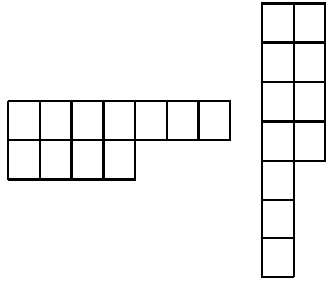
\begin{equation*}
\begin{tabular}{cc}
\begin{tabular}{|l|l|l|l|l|l|l|}
\hline
&  &  &  &  &  &  \\ \hline
&  &  &  \\ \cline{1-4}
\end{tabular}
&
\begin{tabular}{|l|l|}
\hline
&  \\ \hline
&  \\ \hline
&  \\ \hline
&  \\ \hline
  \\ \cline{1-1}
  \\ \cline{1-1}
  \\ \cline{1-1}
\end{tabular}%
\end{tabular}%
\end{equation*}%
The most remarkable feature of the Young patterns as applied to electronic
wave functions is that they are in a one-to-one correspondence with the
total spin, namely: the length of the one-column part of the spatial Young
pattern equals to $2S$. This allowed F.A. Matsen Ref.~ \cite{Matsen:64} yet many
years ago to suggest to avoid any remark concerning the spin in
(nonrelativistic) quantum chemistry context and to replace it by referencing
to the permutational symmetry of the corresponding states. Although, it is,
of course, a matter of terminology, within such a formulation the triplet
component with $S_{z}=+1$ would not ever arise by this hindering any
possible confusion.

The same tools can be used to describe the irreducible representations of
the group of unitary matrices. The \ corresponding construct evolves as
follows Ref. \cite{McWeenybook2nd}: for any number of spatial orbitals $M$
the group $U(M)$ of the unitary $M\times M$ matrices acts as a "dynamical"
group by transforming orbitals. Any given number of electrons $N$ and any
value of the total spin $S$ conforming with two previous values produces an
irreducible representation $\Upsilon $ of the group $U(M)$. As in the case
of the $S_{N}$ group irreducible representations of the group $U(M)$ are
labelled by the Young patterns, but the meaning of their elements is
different. The representation by $N$-electron spatial functions has the
tensor rank $N$ and the corresponding Young pattern contains $N$ boxes
arranged in no more than two columns each of the heigth not larger than $M$,
such that the first column is by $2S$ boxes longer than the second one. This
irreducible representation is degenerate and its rows $\upsilon $ can be
numbered by distributing $M$ orbital symbols in the above $N$ \ boxes in
such a way that they do not decrease (some ordering is assumed among them)
along the rows and strictly increase in each column. Under this rule some
orbital symbols in principle may appear no more than twice in a two-column
pattern by this representing a doubly occupied spatial orbital, those
appearing once represent singly occupied orbitals. Thus constructed Young
tableaux represent states transforming according to the rows $\upsilon $ of
the representation $\Upsilon $. The Young tableau characterizes first of all
the permutation symmetry of the state in that sense that the spatial part of
the many electron function described by the Young tableau $\Upsilon \upsilon
$ is derived from the product of orbitals where each enters as many times as
it appears in the tableau by applying the symmetrization over rows of the
tableau and antisymmetrization over its columns. This construct is also
known as \textit{immanant} wave functions Ref. \cite{Poshusta:68}.

\newpage
\bibliography{myrefs1_iso9,myrefs2_iso9,allold_iso9,reprints_iso9}

\newpage
\begin{table}[tbp]
\label{Tab:Fe_d_shell_energy} 
\center{
\caption{The energy expressions of the many-electron states in the $d$-shell of the Fe$^{2+}$ ion.
$E_0$ is given by the expression: $ E_0 = n_d T + n_d(n_d - 1)A/2  $ where $T$ is the
kinetic energy per electron, $A = F_{0} - 49 F_{4}$ and $n_d$ is the number of electrons in the $d$-shell.}
\begin{tabular}{|rrrrrr|}
\hline
 &&&&& \\
$E(_{\pm }^{1}S)$ & $=$ & $E_0$ & $+10F_{2}$ & $+6F_{4}$ & $\pm \frac{1}{2}\sqrt{3088F_{2}^{2}-26400F_{2}F_{4}+133200F_{4}^{2}}$ \\
$E(_{\pm }^{1}D)$ & $=$ & $E_0$ & $+9F_{2}$ & $-76.5F_{4}$ & $\pm \frac{1}{2}\sqrt{1296F_{2}^{2}-10440F_{2}F_{4}+30825F_{4}^{2}}$ \\
$E(^{2}F)$ & $=$ & $E_0$ &  & $+48F_{4}$ &  \\
$E(_{\pm }^{1}G)$ & $=$ & $E_0$ & $-5F_{2}$ & $-6.5F_{4}$ & $\pm \frac{1}{2}\sqrt{708F_{2}^{2}-7500F_{2}F_{4}+30825F_{4}^{2}}$ \\
$E(^{1}I)$ & $=$ & $E_0$ & $-15F_{2}$ & $-9F_{4}$ &  \\
$E(_{\pm }^{3}P)$ & $=$ & $E_0$ & $-5F_{2}$ & $-76.5F_{4}$ & $\pm \frac{1}{2}\sqrt{912F_{2}^{2}-9960F_{2}F_{4}+38025F_{4}^{2}}$ \\
$E(^{3}D)$ & $=$ & $E_0$ & $-5F_{2}$ & $-129F_{4}$ &  \\
$E(_{\pm }^{3}F)$ & $=$ & $E_0$ & $-5F_{2}$ & $-76.5F_{4}$ & $\pm \frac{1}{2}\sqrt{612F_{2}^{2}-4860F_{2}F_{4}+20025F_{4}^{2}}$ \\
$E(^{3}G)$ & $=$ & $E_0$ & $-12F_{2}$ & $-94F_{4}$ &  \\
$E(^{3}H)$ & $=$ & $E_0$ & $-17F_{2}$ & $-69F_{4}$ &  \\
$E(^{5}D)$ & $=$ & $E_0$ & $-21F_{2}$ & $-189F_{4}$ & \\
\hline
\end{tabular}}
\end{table}
\begin{table}[tbp]
\label{Tab:Fe_d_shell_energy_values}
\center{
\caption{The excitation energy expressions and their values for the many-electron states in the $d$-shell of
the Fe$^{2+}$ ion.}
$$
\begin{tabular}{|l|l|l|}
\hline

$\Delta E(_{-}^{1}G)$ & $16.F_{2}+182.5F_{4}-\frac{1}{2}\sqrt{708F_{2}^{2}-7500F_{2}F_{4}+30825F_{4}^{2}}$ & 30886.4 \\
$\Delta E(_{+}^{1}G)$ & $16.F_{2}+182.5F_{4}+\frac{1}{2}\sqrt{708F_{2}^{2}-7500F_{2}F_{4}+30825F_{4}^{2}}$ & 57221.7 \\
$\Delta E(^{1}I)$ & $6.F_{2}+180.F_{4}$ & 30356.2 \\
$\Delta E(_{-}^{3}P)$ & $16.F_{2}+112.5F_{4}-\frac{1}{2}\sqrt{912F_{2}^{2}-9960F_{2}F_{4}+38025F_{4}^{2}}$ & 20688.4 \\
$\Delta E(_{+}^{3}P)$ & $16.F_{2}+112.5F_{4}+\frac{1}{2}\sqrt{912F_{2}^{2}-9960F_{2}F_{4}+38025F_{4}^{2}}$ & 49576.9 \\
$\Delta E(^{3}D)$ & $16.F_{2}+60.F_{4}$ & 30725.8 \\
$\Delta E(_{-}^{3}F)$ & $16.F_{2}+112.5F_{4}-\frac{1}{2}\sqrt{612F_{2}^{2}-4860F_{2}F_{4}+20025F_{4}^{2}}$ & 21699.9 \\
$\Delta E(_{+}^{3}F)$ & $16.F_{2}+112.5F_{4}+\frac{1}{2}\sqrt{612F_{2}^{2}-4860F_{2}F_{4}+20025F_{4}^{2}}$ & 50276.1 \\
$\Delta E(^{3}G)$ & $9.F_{2}+95.F_{4}$ & 24940.9 \\
$\Delta E(^{3}H)$ & $4.F_{2}+120.F_{4}$ & 20300.8  \\
\hline
\end{tabular}
$$
}
\end{table}

\begin{table}[tbp]
\label{Tab:AverageMultipletEnergies}
\center{
\caption{The average multiplet energies in the $d$-shell of the Fe$^{2+}$ ion.}
\begin{tabular}{|lc|}
\hline
$E(_{\pm }^{1}G)_{\mathrm{av}}$ & 44054. \\
\hline
$E(^{1}I)$ & 30356. \\
\hline
$E(_{\pm }^{3}P)_{\mathrm{av}}$ & 35133. \\
\hline
$E(^{3}D)$ & 30726. \\
\hline
$E(_{\pm }^{3}F)_{\mathrm{av}}$ & 35988. \\
\hline
$E(^{3}G)$ & 24941. \\
\hline
$E(^{3}H)$ & 20301. \\
\hline
\end{tabular}}
\end{table}

\begin{table}[tbp]
\label{Tab:MultipletSplittings}
\center{
\caption{The splittings of the multiplets with coinciding $L$ and $S$ in the $d$-shell of the Fe$^{2+}$ ion.
All values in cm$^{-1}$.}
\begin{tabular}{|l|l|l|l| }
\hline
& calc/noncorr & calc/corr  & exp \\
\hline
$\Delta \left( _{\pm }^{1}G\right) $ & $24140$ & $25984$ & $26335.3$ \\
\hline
$\Delta \left( _{\pm }^{3}P\right) $ & $25993$ & $28255$ & $28888.5$ \\
\hline
$\Delta \left( _{\pm }^{3}F\right) $ & $26142$ & $27726$ & $28576.2$ \\
\hline
\end{tabular}}
\end{table}

\end{document}